\documentclass[10pt,conference]{IEEEtran}
\IEEEoverridecommandlockouts
\usepackage{cite}
\usepackage{amsmath,amssymb,amsfonts}
\usepackage{graphicx}
\usepackage{textcomp}
\usepackage{xcolor}
\usepackage{utfsym}
\usepackage{pifont}
\usepackage{multirow}
\usepackage{tcolorbox}
\usepackage[linesnumbered,ruled,vlined]{algorithm2e}
\usepackage{enumitem}
\usepackage{svg}
\usepackage{subfigure}
\usepackage[doipre={doi:~}]{uri}
\usepackage{booktabs}
\usepackage{bm}
\usepackage{caption}
\usepackage{marvosym}
\usepackage{hyperref}
\def\eg{\emph{e.g.}}
\def\ie{\emph{i.e.}}

\newcommand{\modify}[1]{\textcolor{black}{#1}}

\newcommand{\tool}{\emph{BDefects4NN}\space}
\newcommand{\toolns}{\emph{BDefects4NN}}
\newcommand{\slicetool}{SLICER\space}

\newcommand{\totalnumber}{1,654\space}

\newcommand{\revised}[1]{\textcolor{black}{#1}}

\def\BibTeX{{\rm B\kern-.05em{\sc i\kern-.025em b}\kern-.08em
    T\kern-.1667em\lower.7ex\hbox{E}\kern-.125emX}}
\begin{document}

\title{ \textit{BDefects4NN}: A Backdoor Defect Database for Controlled Localization Studies in Neural Networks}

\author{
\IEEEauthorblockN{Yisong Xiao$^{1,2}$,
Aishan Liu$^{1}$\textsuperscript{\Letter},%\IEEEauthorrefmark{1},
Xinwei Zhang$^{1}$,
Tianyuan Zhang$^{1,2}$,
Tianlin Li$^{3}$,\\
Siyuan Liang$^{4}$,
Xianglong Liu$^{1,5}$,
Yang Liu$^{3}$,
Dacheng Tao$^{3}$}
\IEEEauthorblockA{$^{1}$SKLCCSE, Beihang University, Beijing, China \quad$^{2}$Shen Yuan Honors College, Beihang University, Beijing, China \\
\quad$^{3}$Nanyang Technological University, Singapore \\
$^{4}$National University of Singapore, Singapore \quad$^{5}$Zhongguancun Laboratory, Beijing, China}
}

\maketitle

\begin{abstract}

Pre-trained large deep learning models are now serving as the dominant component for downstream middleware users and have revolutionized the learning paradigm, replacing the traditional approach of training from scratch locally. To reduce development costs, developers often integrate third-party pre-trained deep neural networks (DNNs) into their intelligent software systems. However, utilizing untrusted DNNs presents significant security risks, as these models may contain intentional backdoor defects resulting from the black-box training process. 
These backdoor defects can be activated by hidden triggers, allowing attackers to maliciously control the model and compromise the overall reliability of the intelligent software. To ensure the safe adoption of DNNs in critical software systems, it is crucial to establish a backdoor defect database for localization studies. This paper addresses this research gap by introducing \toolns, the first backdoor defect database, which provides labeled backdoor-defected DNNs at the neuron granularity and enables controlled localization studies of defect root causes. 

In \toolns, we define three defect injection rules and employ four representative backdoor attacks across four popular network architectures and three widely adopted datasets, yielding a comprehensive database of 1,654 backdoor-defected DNNs with four defect quantities and varying infected neurons. Based on \toolns, we conduct extensive experiments on evaluating six fault localization criteria and two defect repair techniques, which show limited effectiveness for backdoor defects. Additionally, we investigate backdoor-defected models in practical scenarios, specifically in lane detection for autonomous driving and large language models (LLMs), revealing potential threats and highlighting current limitations in precise defect localization. This paper aims to raise awareness of the threats brought by backdoor defects in our community and inspire future advancements in fault localization methods.

\end{abstract}

\begin{IEEEkeywords}
Backdoor defects, fault localization, deep learning
\end{IEEEkeywords}

\section{Introduction}
\vspace{-0.05in}

Deep Learning (DL) has demonstrated remarkable performance across a wide range of applications and is integrated into diverse software systems, such as autonomous driving \cite{bojarski2016end} and healthcare \cite{janowczyk2016deep,miotto2018deep}. 
A consensus is emerging among developers to employ DL models pre-trained on large-scale datasets for their downstream applications \cite{han2021pre,niu2022spt,alshalali2018fine,zhang2020sentiment,robbes2019leveraging}. By fine-tuning publicly available pre-trained model weights on their specific datasets, developers with limited resources or training data can effortlessly craft high-quality models for a multitude of tasks. As a result, the pre-training and fine-tuning paradigm has gained strong popularity \cite{ding2023parameter}.

However, these third-party released DNNs are often pre-trained on large-scale, noisy, and uncurated Internet data that are unknown to downstream users. Utilizing these untrusted DNNs presents significant safety risks, as these models may contain intentional \emph{backdoor defects} resulting from the black-box training process (\revised{we refer to models with backdoor defects as ``backdoor-defected models'', also ``infected models'' for convenience}). These malicious defects in DNN models are caused by backdoor attacks \cite{li2022backdoor}, where an attacker adversarially injects backdoored neurons into the victim models by poison training or \modify{sub-network replacing}, thereby being able to manipulate the model behavior with a specific trigger.  For example, a third-party released DNN that is injected with backdoor defects will incorrectly identify lanes triggered by two common traffic cones when deployed into autonomous driving systems \cite{han2022physical}, thereby compromising the overall reliability of the intelligent systems and endangering human lives. To ensure the safe adoption of DNNs in critical software systems, it is crucial to establish a comprehensive \modify{backdoor defect database} for \modify{localization studies}. However, current defect localization \modify{databases} for DNNs \cite{wardat2021deeplocalize,wardat2022deepdiagnosis,ghanbari2023mutation} primarily focus on \emph{common defects}, \ie, unintentional functional bugs introduced by DNN developers such as incorrect tensor shapes, while overlooking the stealthy and harmful defects posed by backdoor attacks (\ie, \modify{backdoor defects}). This sparsity of research presents severe safety risks to DL systems.
% as it increases its vulnerabilities to backdoor defects.
% \footnote{We use ``backdoor defects'' or ``injected faults'' interchangeably.}
% \footnote{\revised{Models with backdoor defects are also referred to as "infected models."}}

% To bridge the gap, this paper takes the first step in comprehensively studying and benchmarking the backdoor defect localization in DNNs. 
To bridge the gap, this paper takes the first step in constructing a backdoor defect database for localization studies in DNNs.
We propose the first comprehensive backdoor defect database \toolns, which provides neuron-level backdoor infected DNNs with ground-truth defect labeling to support defect localization studies at the neuron granularity, \revised{serving as an essential test suite for our community}. Specifically, we propose the defect design protocols to select neurons for defect injection in terms of neuron contribution, neuron quantity, and sub-network correlation. Based on the proposed defects design protocols, we employ four representative backdoor attack methods to inject backdoor defects into four popular network architectures across three widely adopted datasets on the image classification task. Overall, our \tool contains 1,654 backdoor-defected DNNs with ground truth defects labeling, categorized into 48 directories, each featuring injected sub-networks at \revised{four (4)} quantity levels, offering varying defect quantities and infected neurons \revised{to enable comprehensive evaluation of localization and repair methods}. 

Using our \toolns, we conduct extensive experiments to evaluate the performance of six fault localization criteria, incorporating four backdoor-specific criteria from the backdoor defense domain and two general criteria from the software engineering field. Notably, we identify that current fault localization methods show limited performance on backdoor defects in DNNs with low localization effectiveness (17.64\% $WJI$ on average). In addition, we further evaluate two defect repair techniques, namely neuron pruning and neuron fine-tuning, which exhibit an average $ASRD$ of 39.54\% and 41.45\%, respectively.
Furthermore, we extend our investigations to practical scenarios, such as lane detection (LaneATT \cite{tabelini2021keep}) for autonomous driving and LLMs (ChatGLM \cite{du2022glm}), where we illustrate the potential threats posed by backdoor defects and highlight the current limitations of existing methods in precisely localizing these defects in real-world applications.
We hope this paper will raise awareness of backdoor defect threats within our community and facilitate further research on fault localization methods. Our main \textbf{contributions} are:

\begin{itemize} [leftmargin=*]
% \vspace{-0.1in}
\item 
As far as we know, we pioneer the integration of backdoor defects into the fault localization task and conduct the first comprehensive study on backdoor defect localization in DNNs.

\item We build \toolns, a comprehensive database containing \totalnumber backdoor-defected DNNs with neuron-level ground-truth labeling, \modify{supporting controlled defect localization studies}.

\item We conduct extensive evaluations on six localization criteria and two defect repair methods, offering findings into their strengths and weaknesses.

\item We publish \tool as a self-contained toolkit on \revised{our website} \cite{ourweb}.

\end{itemize}

% \clearpage
\vspace{-0.025in}
\section{Preliminaries}
% \vspace{-0.05in}

\textbf{\emph{DNN}}. 
Given a dataset $\bm{D}$ with data sample $\bm{x} \in X$ and label $y \in Y$, the deep supervised learning model aims to learn a mapping or classification function $F_{\Theta}: X \rightarrow Y$. The model $F_{\Theta}$ consists of $L$ serial layers, with parameters $\Theta = \{\bm{\theta}_{1},...,\bm{\theta}_{L}\}$, and $N_{l}$ neurons in each layer $l \in \{ 1, ..., L \}$.  The total number of neurons is $N = \sum_{l=1}^{L} N_{l}$. 
Further, we denote the activation output of each neuron $F_{l}^{i}$ as $\bm{a}_{l}^{i}$, where $i \in \{1,...,N_{l}\}$.
Moreover, a sub-network is defined as a pathway within $F_{\Theta}$ that includes at least one neuron in each layer $l$, where $l \in \{ 1, ..., L-1 \}$.
This paper mainly focuses on the image classification task, with its training process as:
\begin{equation}
    \Theta = \arg \min_{\Theta} \mathbb{E}_{(\bm{x}, {y}) \sim \bm{D}}[\mathcal{L}(F_{\Theta}(\bm{x}), {y})],
\end{equation}
where $\mathcal{L}(\cdot)$ represents the cross-entropy loss function.

\textbf{\emph{Backdoor Attack}}.

Backdoor attacks aim to embed hidden behaviors into a DNN $F_{\Theta}$ during training, allowing the infected model $F_{\hat{\Theta}}$ to behave normally on benign samples. However, the predictions of the infected model undergo malicious and consistent changes when hidden backdoors are activated by attacker-specified trigger patterns. Presently, poisoning-based backdoor attacks stand as the most straightforward and widely adopted method in the training phase. Specifically, the attacker randomly selects a small portion $p$ (\eg, 10\%) of clean data from the training dataset $\bm{D}$, and then generates poisoned samples $\hat{\bm{D}} =\{(\hat{\bm{x}}_i, \hat{y}_i)\}_{i=1}^M, M = p \cdot \vert \bm{D} \vert$, by applying the trigger $\bm{T}$ to the images using the function $\phi$ and modifying the corresponding label to the target label $\hat{y}_i$ as follows:
\begin{equation}
\label{eqn:backdoor_generation}
    \hat{\bm{x}}_{i} = \phi (\bm{x}_{i}, \bm{T}) , \quad \hat{{y}}_i = \eta(y_i).
\end{equation}
For different backdoor attack methods, the trigger generation function $\phi$ varies, and $\eta$ represents the rules governing the modification of poisoning labels. 
Afterward, the model trained on the poisoned dataset $\bm{D} \cup \hat{\bm{D}}$ will be injected with backdoors, yielding target label predictions $F_{\hat{\Theta}}(\hat{\bm{x}}_{i}) = \hat{{y}_i}$ on test images $\hat{\bm{x}}_{i}$ containing triggers. 
% Besides poisoning-based backdoor attacks, 
Another series \revised{(structure-modified attacks)} \cite{qi2022towards,qi2021subnet,tang2020embarrassingly,li2021deeppayload} first trains a backdoor sub-network/module, and then directly injects the sub-network into a benign model $\Theta$ to obtain the final infected model $\hat{\Theta}$. 
\revised{Backdoor training involves dual-task learning: the clean task on clean dataset $\bm{D}$ and the backdoor task on backdoor dataset $\hat{\bm{D}}$.}
\revised{An infected model should achieve high attack success rate $ASR$ (backdoor task) and competitive clean accuracy $CA$ (clean task):
\begin{align}\begin{aligned}
CA &= P_{(\bm{x},y)\sim \bm{D}_{test}}{(F_{\hat{\Theta}}(\bm{x})=y)}, \\
ASR &= P_{(\hat{\bm{x}},\hat{y})\sim \hat{\bm{D}}_{test}}{(F_{\hat{\Theta}}(\hat{\bm{x}})=\hat{y})},
\end{aligned}\end{align}
where $\bm{D}_{test}$ and $\hat{\bm{D}}_{test}$ denote the clean test dataset and poisoned test dataset respectively.
}

\revised{\textbf{\emph{Assumption}}. We further state the \textbf{\emph{common assumption}} in infected DNNs: \textbf{specific neurons/sub-nets are predominantly responsible for backdoor defects}, which has been widely accepted and empirically demonstrated in backdoor attack and defense studies \cite{gu2017badnets,liu2018trojaning,qi2022towards,qi2021subnet,tang2020embarrassingly,li2021deeppayload,liu2018fine,wang2019neural,wu2021adversarial,zheng2022data,guan2024backdoor,sun2024neural}. From the perspective of dual-task learning, numerous studies \cite{li2023reconstructive,li2021neural} have revealed the fact that neurons in infected models can be decomposed into clean and backdoor neurons since backdoor attacks are designed not to impact the model’s performance on clean samples \cite{gu2017badnets}, indicating a high level of independence between the clean and backdoor tasks \cite{li2023reconstructive}. In poisoning attacks, TrojanNN \cite{liu2018trojaning} optimizes triggers to maximize the activation of a few specific neurons for backdoor behavior, while most neurons continue to perform normal functions. In structure-modified attacks \cite{qi2022towards,qi2021subnet,tang2020embarrassingly,li2021deeppayload}, the injected sub-network is responsible for backdoor defects. Furthermore, defense studies design rules based on activation \cite{liu2018fine,wang2019neural}, weight \cite{wu2021adversarial,zheng2022data}, and Shapley value \cite{guan2024backdoor} to identify and repair the neurons most responsible for backdoor, thereby eliminating the backdoor. Thus, following common assumptions, we aim to inject backdoor defects into DNNs at specific neurons/sub-nets, providing defect labeling to support localization and repair studies.}
\section{\tool Database}
\vspace{-0.025in}
\begin{figure*}[!ht]
\centering
\includegraphics[width=0.9\linewidth]{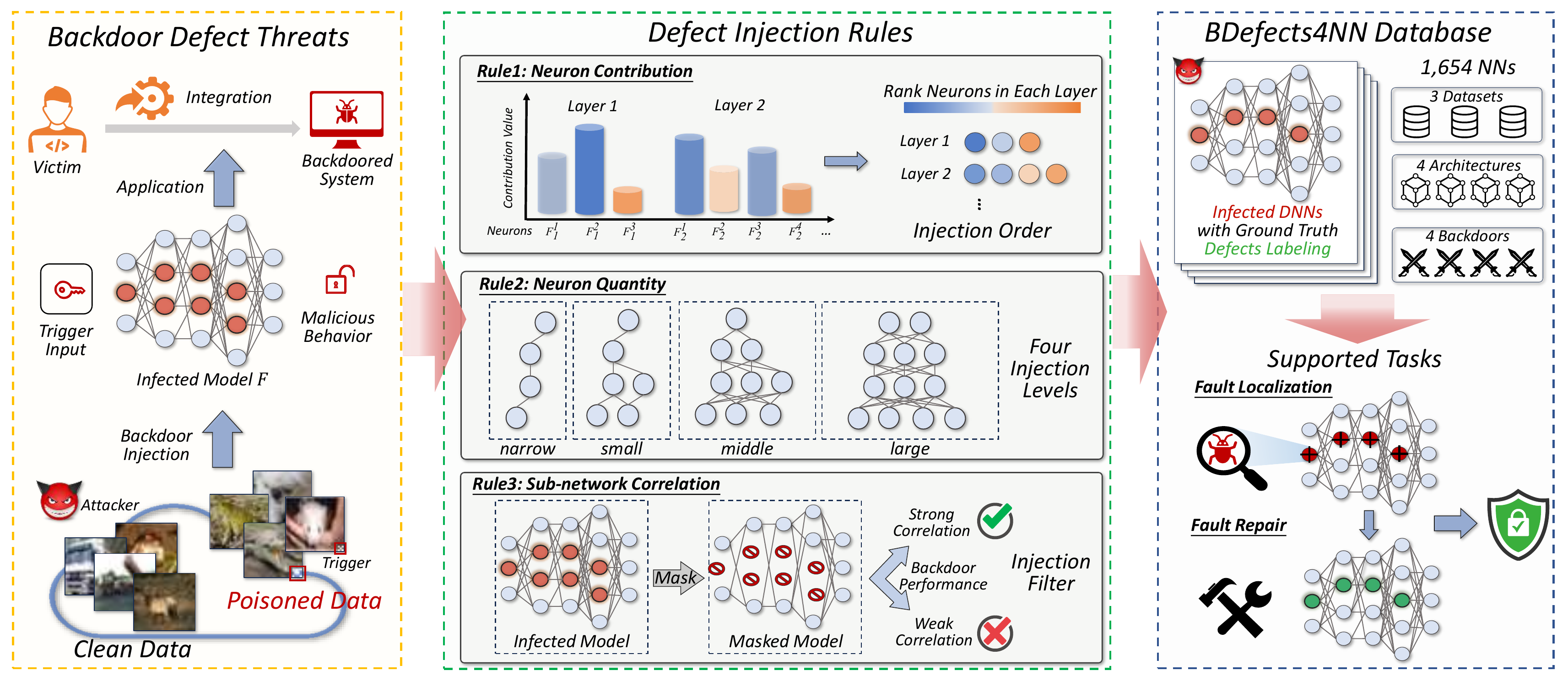}
\vspace{-0.05in}
\caption{Overview of \tool framework. Targeting image classification task, our \tool designs three rules to inject neuron-level backdoors into DNNs and builds 1,654 DNNs with backdoor defects, which can support the evaluation of fault localization methods and defect repair techniques.
}
\label{fig:framework}
\vspace{-0.2in}
\end{figure*}

In this section, we first illustrate the motivation and problem definition, then explain the \tool design protocol and construction details. Figure \ref{fig:framework} shows the overall framework.

\subsection{Motivations}

\textbf{\emph{Possible Threat Scenarios.}} Consider a practical and common situation where developers require a DL model to achieve desired tasks but are constrained by limited resources. In such cases, they might resort to using a third-party platform (\eg, cloud computing platforms) for training or opt to download and utilize a pre-trained DNN model directly provided by a third party \cite{li2022backdoor}. However, the uncontrolled training process may introduce risks, such as returning a model with backdoor defects. Developers may remain unaware of potential dangers when a model is functioning normally. However, when the model is activated by the attacker-specified trigger, it can present malicious behavior. 
For instance, an infected lane detection model within the autonomous driving system may cause the vehicle to deviate from the road when encountering two traffic cones \cite{han2022physical}, which will be further discussed in Section \ref{sec:case_study_lane_detection}.
Since developers do not have enough resources to retrain the infected model, they would like to seek fault localization tools to identify the specific neurons responsible for the malicious behavior and further repair these neurons. This parallels how developers encountering bugs in DL programs utilize existing fault localization methods to identify and address the bugs \cite{wardat2022deepdiagnosis,wardat2021deeplocalize}.

\textbf{\emph{Problem Definition.}} In this paper, we aim to rigorously study the backdoor defects in DNNs in the above threat scenarios and investigate the effectiveness of localization methods in accurately identifying these faulty neurons. Formally, the research problem can be represented as follows: given a DNN $F_{\hat{\Theta}}$, \modify{we inject it with infected neurons} $S^{fault}$, and a localization method is employed to identify the suspicious neurons $S^{localized}$; how closely aligned are these two sets (\ie, $S^{fault}$ and $S^{localized}$); and what is the model performance after repairing on $S^{localized}$. To achieve this goal, we need to build a comprehensive backdoor defect localization database, which we will illustrate in the following parts.

\subsection{Backdoor Defects Design Protocol}

To build a comprehensive backdoor defect database, we propose backdoor defect design protocols. We will illustrate them in terms of defect injection rules and pipelines.

\textbf{Defect Injection Rules.} In the image classification task, a convolutional neural network (CNN) model is composed of multiple layers, each housing numerous kernels (\ie, neurons). Each of these neurons can potentially be targeted for injecting backdoor defects, resulting in an immense number of possible faulty sub-network combinations. For instance, in a 10-class VGG-16 network with 4,224 kernels, there are approximately an overwhelming $2^{4,224}$ potential sub-network candidates. Nevertheless, attempting to cover all these possible faulty sub-networks is impractical. Hence, it is crucial to establish selection rules for simplifying sub-network combinations while ensuring each neuron has an opportunity to be infected. Specifically, our selection rules for sub-networks are based on \textbf{neuron contribution}, \textbf{neuron quantity}, and \textbf{sub-network correlation}.

\textbf{Rule} \ding{182}: \emph{Neuron Contribution}. Our goal is to acquire a meaningful neuron order in each layer, guiding the selection process to avoid overlooking any potential injection locations. Inspired by neural network interpretation methods \cite{khakzar2021neural,molchanov2019importance}, we adopt NeuronMCT \cite{khakzar2021neural} to calculate the neuron contribution to the model predictions, which quantifies the influence degrees of a specific neuron on the overall model behavior. Specifically, for neuron $F_{l}^{i}$ with output $\bm{a}_{l}^{i}$, its neuron contribution $c_{l}^{i}$ can be calculated as follows:
\begin{equation}
\label{eqn:neuronmct}
    c_{l}^{i} = \vert F_{\Theta}(\bm{x}) - F_{\Theta}(\bm{x},\bm{a}_{l}^{i} \leftarrow 0) \vert  \simeq \vert \bm{a}_{l}^{i} \nabla_{\bm{a}_{l}^{i}} F_{\Theta}(\bm{x}) \vert,
\end{equation}
where a Taylor approximation is employed to achieve faster calculations. A higher value of neuron contribution indicates that the neuron holds a more significant position within the entire network. 
For neurons in layer $l$, their contributions are:
\begin{equation}
    C_{\bm{\theta}_{l}} = \{ c_{l}^{i} |  1 \leq i \leq N_{l} \} .
\end{equation}

Therefore, we could rank neurons in layer $l$ based on their contributions $C_{\bm{\theta}_{l}}$. And we denote the neuron contribution order as $\Pi_{\bm{\theta}_{l}}$, which access the original neuron $F_{l}^{i}$ via a reverse map of rank order as:
\begin{equation}
    \Pi_{\bm{\theta}_{l}} = \{ \pi_{l}({j}) |  1 \leq j \leq N_{l} \},
\end{equation}
where $\pi_{l}({j})$ represents the original index of neuron with $j$-th contribution. For example, $F_{l}^{\pi_{l}({1})}$ is the neuron with the highest contribution. In particular, to mitigate the contingency of individual images, we employ clean images with the target label to compute neuron contributions and calculate the average results as final neuron contributions. After obtaining the rank of neuron contributions, we utilize $\Pi_{\bm{\theta}_{l}}$ of each layer as subsequent neuron selection guidance.

\textbf{Rule} \ding{183}: \emph{Neuron Quantity}. Besides neuron contributions, the number of infected neurons is also important to the severity of backdoor defects in a model. Therefore, we aim to inject defects using different numbers of neurons, which simulates \revised{defects of varying sizes} and simplifies the formation of sub-network combinations. Inspired by group schemes designed to reduce extensive search space \cite{qi2022patching}, we empirically set four levels in the neuron number to group sub-networks, including \emph{narrow}, \emph{small}, \emph{middle}, and \emph{large}. 
Specifically, for the \emph{narrow} level, we keep alignment with SRA \cite{qi2022towards}, which selects one or two neurons in each layer (the exact number is determined by network architectures and layers); for the \emph{small}, \emph{middle}, and \emph{large} levels, we choose neurons in each layer based on a specified percentage: 5\% for \emph{small}, 10\% for \emph{middle}, and 20\% for \emph{large}. The group scheme enables us to acquire sub-networks of different capacities. \revised{Note that} the top 1\% (whose number is almost equivalent to the \emph{narrow} level) of backdoor-related neurons is adequate to activate the backdoor behavior \cite{wang2019neural} \revised{since the backdoor task is much easier than the clean task}. 
\revised{The 20\% for \emph{large} level is based on that removing 20\% of suspicious neurons nearly eliminates backdoor behavior \cite{wang2019neural}.}

\textbf{Rule} \ding{184}: \emph{Sub-network Correlation}. However, there may exist a disparity between selected sub-networks and backdoor attacking performance \revised{($ASR$)}. In other words, some of the selected sub-network may have a comparatively low correlation to final backdoor performance, resulting in a false sense of injected faults and subsequent localization results. To mitigate this, we tailor the sub-networks that have high correlations to the model backdoor effects by calculating the correlation rate  \cite{xie2022npc}. 
Specifically, to measure whether the injected sub-network has a critical impact on the backdoor prediction of the infected model, we follow \revised{NPC} \cite{xie2022npc} mask the outputs of the injected sub-network as zero and utilize the drop rate on \revised{$ASR$} as its correlation rate, denoted as $ASR.Cor$: 
\begin{equation}
ASR.Cor = ASR_{\hat{\Theta}} - ASR_{\hat{\Theta}_{m}},
\end{equation}
where $\hat{\Theta}_{m}$ is the masked model. \revised{The masking test is conducted after backdoor injection}, and a higher $ASR.Cor$ indicates a larger impact of sub-networks on backdoor predictions. \revised{Empirically, we retain infected models with $ASR.Cor > 0.5$ due to its polarized correlation distribution, which effectively distinguishes neurons primarily responsible for backdoor task.}

\textbf{Defect Injection Pipeline.} Based on the above rules, we can treat each layer $l$ as a sorted list (\ie, neuron order $\Pi_{\bm{\theta}_{l}}$); we sequentially choose neurons (without replacement) within each layer based on their contribution order, stopping when the desired quantity is reached (determined by the layer's neuron count $N_{l}$ and the specified quantity level). The neurons selected in this process form a sub-network denoted as $S^{fault}$, and we inject backdoor defects into this sub-network to obtain an infected model $\hat{\Theta}$. We then calculate the \revised{$ASR.Cor$} of $\hat{\Theta}$ after masking $S^{fault}$, and retain $\hat{\Theta}$ in our database if its \revised{$ASR.Cor$} exceeds 0.5.
This process is iterated until all neurons have been accounted for, \revised{achieving coverage from high-impact to low-impact neurons for injection}. 

Taking the \emph{small} level (5\%) as an example, we can formulate its sub-network selection process as follows: 
\begin{align}\begin{aligned}
    S^{fault} = & \{ F_{l}^{\pi_{l}({j})} |  1 \leq l \leq L-1, \\
    & 1 + \lfloor 5\% \cdot i \cdot N_{l} \rfloor \leq j \leq \lfloor 5\% \cdot (i+1) \cdot N_{l} \rfloor \}, 
\end{aligned}\end{align}
where $i$ represents the $i$-th selection, for the \emph{small} level, a total of 20 selections are made. For the \emph{middle} and \emph{large} levels, the selection process is similar, with replacement percentages modified to 10\% and 20\%, respectively. As for the \emph{narrow} level, we follow the same 20 selections as the \emph{small} level to reduce sub-network quantities, with the distinction that it retains only the first one or two neurons in each layer selection. As a result, we generate a total of 55 sub-network candidates for injection, including 20 \emph{narrow}, 20 \emph{small}, 10 \emph{middle}, and \revised{five (5)} \emph{large} sub-network candidates. Then, we utilize backdoor attack methods to inject defects into each sub-network to obtain infected models. Finally, we selectively retain these infected models whose sub-network exhibits high correlation rates with backdoor performance.

In addition, since infected neurons may have varying roles in the backdoor response, we offer the \textbf{weight} of each infected neuron to backdoor performance as affiliate information, where we use NeuronMCT to calculate their contributions to the model response when triggers are fed, aiming to achieve a more precise assessment of subsequent localization. The weight is denoted as $RC$:
\begin{equation}
    RC = \{ \frac{c_i }{\sum_{i=1}^{m}{c_i}} |  1 \leq i \leq  m \}, \quad m = \vert S^{fault} \vert
\end{equation}
where $c_i$ is the contribution of neuron $S^{fault}_i$.

\revised{Algorithm \ref{algo_all} illustrates the overall process of our  \textbf{defect injection pipeline}, involving the following steps}: first, we generate sub-networks as potential subjects for defect injection (the first two injection rules); \revised{then, we apply attack methods on each sub-network to obtain backdoor-defected models and selectively retain those sub-networks with high correlation to the backdoor performance (the last injection rule)}; finally, we assign infected neurons with their relative contributions on backdoor effects for more precise localization evaluation. \revised{By injecting selected sub-networks, our pipeline enables standardized and fair comparisons of localization and repair methods across diverse attacks.}

\begin{algorithm}[ht]
\footnotesize
\caption{\revised{defect injection pipeline}}
\label{algo_all} 
    
\SetKwInOut{Input}{Input}
\SetKwInOut{Output}{Output}
\SetKwComment{Comment}{/* }{ */}

\revised{
\Input{Benign model $\Theta$, layers number $L$, backdoor attack $\mathcal{A}$.} %, clean dataset $\bm{D}$, backdoor dataset $\hat{\bm{D}}$ 
\Output{A set of defect-labeled infected models $\mathcal{D}_{infected}$.}
\BlankLine 
%%%%% first generate sub-networks
\tcp{generate sub-network candidates}
% $\Pi$ $\leftarrow$ $\{ \Pi_{\bm{\theta}_{l}} | 1 \leq l \leq L-1 \}$\Comment*[r]{acquire neuron order of each layer in benign model $\Theta$ using \textbf{rule 1}}
$\Pi \leftarrow \emptyset$; \tcp{set of each layer's neuron order in $\Theta$}
\For{$l = 1$ \KwTo $L-1$}{
        $\Pi_{\bm{\theta}_{l}} \leftarrow$ acquire layer $l$'s neuron order via \textbf{rule 1}; \tcp{Eq.(6)}
        $\Pi \leftarrow \Pi \cup \Pi_{\bm{\theta}_{l}}$\;
    }
$\mathcal{S}_{candidate} \leftarrow \emptyset$; \tcp{sub-network candidates}
\ForEach{selections $\in\{$20, 20, 10, 5$\}$}{ \tcp{\textbf{rule 2}: narrow,small,middle,large level}
    \For{$i = 0$ \KwTo selections}{
        $S^{fault} \leftarrow$ select sub-network based on $\Pi$ like Eq. (8)\Comment*[r]{replace percentages for middle and large; special deal for narrow level}
        $\mathcal{S}_{candidate} \leftarrow \mathcal{S}_{candidate} \cup S^{fault}$\;
    }
}
\Comment{conduct injection, retain high-correlation sub-networks, and assign contribution}
$\mathcal{D}_{infected} \leftarrow \emptyset$; \tcp{sub-network candidates}
\ForEach{$S^{fault} \in \mathcal{S}_{candidate}$}{
    $\hat{\Theta} \leftarrow $ apply backdoor injection $\mathcal{A}$ on $S^{fault}$\;
    $ASR.Cor \leftarrow $ calculate backdoor correlation via \textbf{rule 3} Eq.(7)\;
    \If{$ASR.Cor>0.5$}{
        $RC \leftarrow $ assign infected neurons' contribution by Eq. (9)\;
        $\mathcal{D}_{infected} \leftarrow \mathcal{D}_{infected} \cup (\hat{\Theta}, S^{fault}, RC)$\;
    }
}
\textbf{return} $\mathcal{D}_{infected}$\;
}
\end{algorithm}
% \DecMargin{1em}
\vspace{-0.1in}

\subsection{Database Construction Details}
\textbf{Backdoor Injection Methods.} In this study, we choose four representative backdoor attack methods to inject defects into DNNs, including BadNets \cite{gu2017badnets}, Blended \cite{chen2017targeted}, TrojanNN \cite{liu2018trojaning}, and SRA \cite{qi2022towards}. 
The first three methods are poisoning-based attacks, where we \revised{inject by fine-tuning the selected sub-network $S^{fault}$ and the classification head of a benign model on the poisoned dataset, restricting weight updates to these components only}. For the structure-modified attack SRA \cite{qi2022towards}, we train an independent sub-network $S^{fault}$ to learn the backdoor, then replace the corresponding sub-network in the benign model, severing its interactions with the rest of the model.
\revised{Following default settings \cite{gu2017badnets,chen2017targeted,liu2018trojaning,qi2022towards}, we employ consistent triggers (\eg, position and distribution), with details available on our website \cite{ourweb}.}

\textbf{Datasets and Models.} Targeting image classification task, \tool utilizes three widely employed datasets in DL and backdoor attack research \cite{wu2022backdoorbench,gu2017badnets,sun2022causality,pang2022trojanzoo}, including CIFAR-10 \cite{krizhevsky2009learning}, CIFAR-100 \cite{krizhevsky2009learning}, and GTSRB \cite{stallkamp2012man}. For models, \tool employs four popular network architectures, including VGG-13 and VGG-16 in VGG series \cite{simonyan2014very}, as well as ResNet-18 and ResNet-34 in ResNet series \cite{he2016deep}. Description of datasets and models can be found on our website \cite{ourweb}.

\begin{figure}[t]
    \vspace{-0.15in}
    \centering
    \subfigure[Defects across three datasets]{
        \includegraphics[width=0.47\linewidth]
        {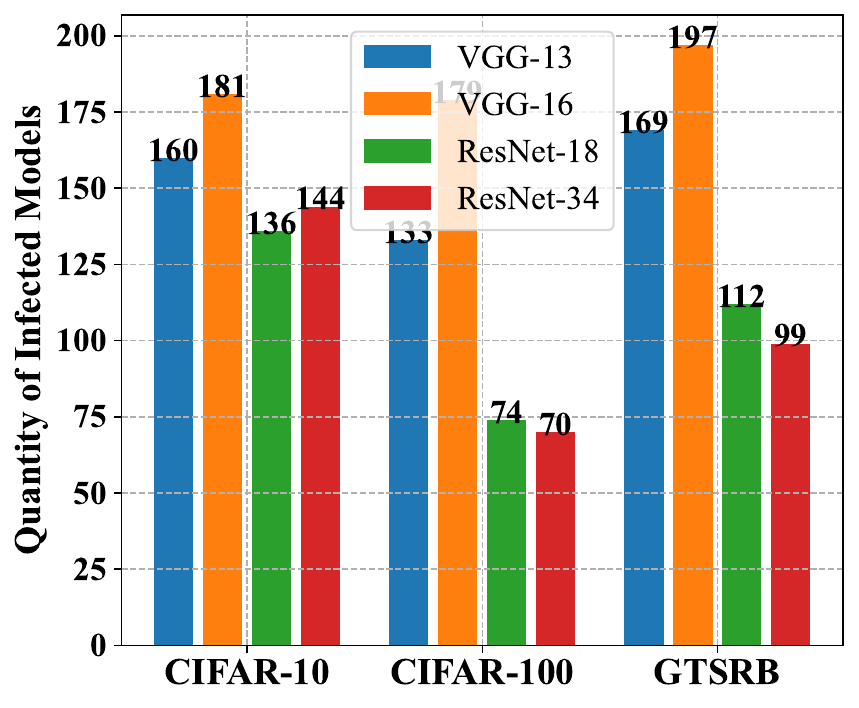}
    }
    \subfigure[Defects on the CIFAR-10 dataset]{
        \includegraphics[width=0.45\linewidth]
        {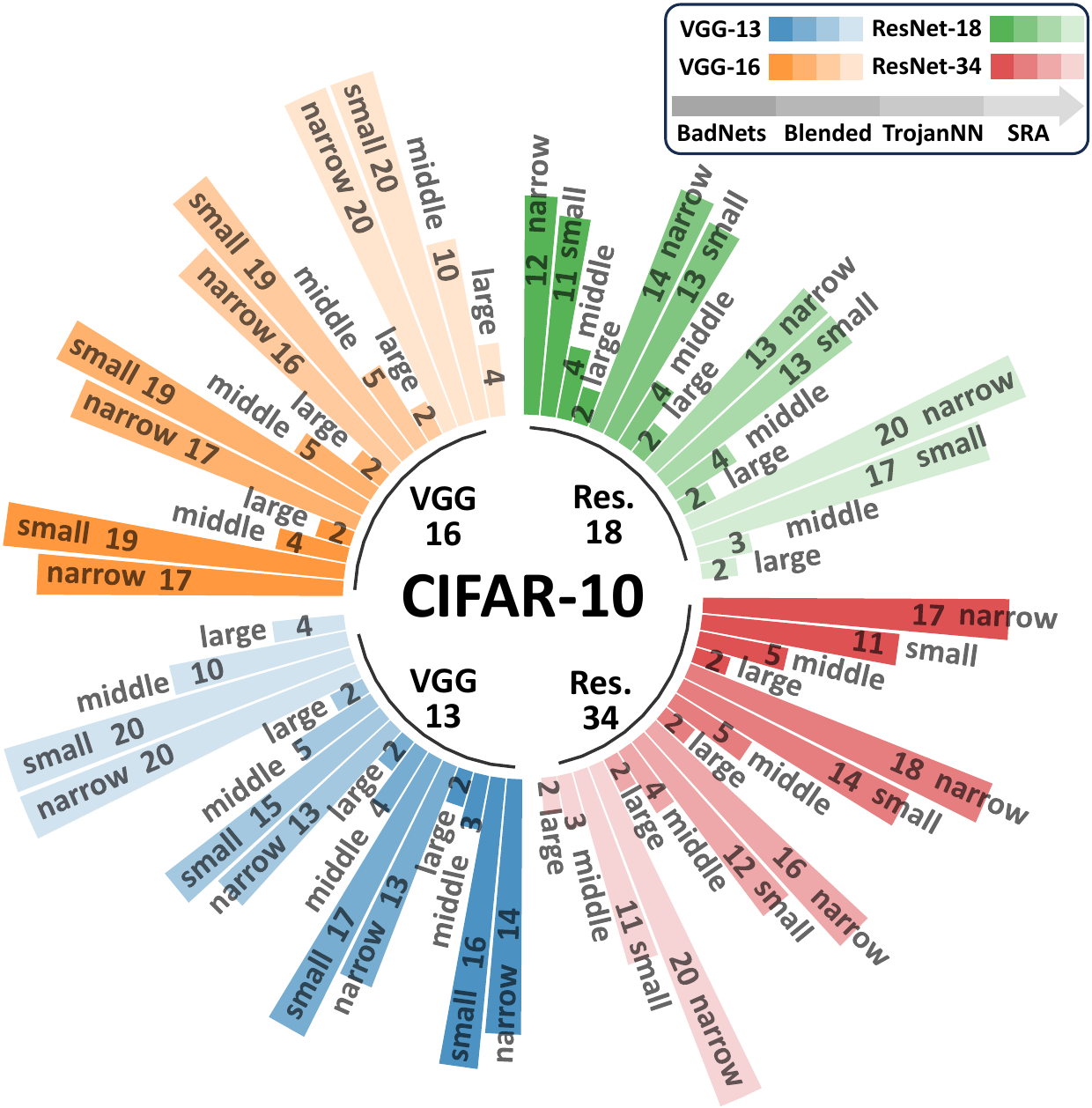}
    }
    \vspace{-0.1in}
    \caption{Defects distribution of \tool database.}
    \vspace{-0.2in}
    \label{fig:database_statistic}
\end{figure}

\vspace{-0.05in}
\subsection{Database Properties}
\vspace{-0.05in}
Across three datasets and four architectures, we generate 55 sub-networks for each DNN and inject defects through \revised{four (4)} attacks into each sub-network. After preserving sub-networks exhibiting high correlation rates, our \tool comprises 1,654 backdoor-infected DNNs with ground truth defect labeling. These are organized into 48 directories, with each directory containing sub-networks at \revised{four (4)} quantity levels. As shown in Figure \ref{fig:database_statistic}, \revised{our defect injection generally proves effective, resulting in 621 (70.57\%) infected models on CIFAR10, 456 (51.82\%) on CIFAR100, and 577 (65.57\%) on GTSRB, with high correlation rates. 
The lower proportion on CIFAR100 can be attributed to its larger number of classes, which reduces neuron redundancy.}

\vspace{-0.05in}
\subsection{Database Usage}

Based on our meticulously constructed database, we can use it to evaluate different tasks as follows.

\textbf{Fault Localization.} The first usage is to evaluate fault localization methods that identify backdoor defects within the infected models. Given an infected DNN $F_{\hat{\Theta}}$ with the defective sub-network $S^{fault}$ and a small portion of clean data as inputs, the fault localization method produces a suspicious sub-network denoted as $S^{localized}$.
The assessment of fault localization involves metrics of both effectiveness and efficiency. In terms of effectiveness, considering a sub-network as a set of neurons, we can utilize the Weighted Jaccard Index ($WJI$) between the sets $S^{fault}$ and $S^{localized}$ as a measure:
% , which can be written as
\begin{equation}
    WJI = \frac{\sum_{i=1}^{m} \sum_{j=1}^{n} \mathbb{I}(S^{fault}_i = S^{localized}_j) \cdot RC_i \cdot \vert S^{fault} \vert }{ \vert S^{fault} \cup S^{localized} \vert},
\end{equation}
where $m = \vert S^{fault} \vert,n = \vert S^{localized} \vert$, $RC$ is the relative contributions of infected neurons, and $\mathbb{I}(\cdot)$ is the indicator function. $\mathbb{I}(A)=1$ if and only if the event ``$A$'' is true.
Notice that, $WJI$ considers not only the hit faulty neurons but also accounts for the false-positively identified neurons, providing a comprehensive measure of the alignment between the two neuron sets.
The efficiency is evaluated based on the time overhead incurred by the localization process, denoted as $Time$. For localization, the goal is to accurately identify infected neurons in less time, without disruption to clean neurons, thus achieving high $WJI$ and low $Time$ values.

\textbf{Fault Repair.} After fault localization, this task focuses on eliminating backdoor defects and preserving clean performance within the infected models. Given an infected DNN $F_{\hat{\Theta}}$, the corresponding suspicious sub-network $S^{localized}$, and a small portion of clean data as inputs, the fault repair process produces a repaired model $F_{\bar{\Theta}}$. 
Evaluation of repair performance involves using clean accuracy drop ($CAD$) and attack success rate drop ($ASRD$), where a successful repair is characterized by a high $ASRD$ and a low $CAD$.
Specifically, $CAD$ and $ASRD$ are calculated \revised{as follows}:
\begin{equation}
    CAD = CA_{\hat{\Theta}} - CA_{\bar{\Theta}}, \quad
    ASRD = ASR_{\hat{\Theta}} - ASR_{\bar{\Theta}}.
\end{equation}

\section{Evaluation}
We first outline the experimental setup and then conduct the evaluation to answer the following research questions. 

\textbf{RQ1}: 
What are the features of the infected neural networks within our \tool database?

\textbf{RQ2}: How effective and efficient are the six localization criteria in localizing backdoor-defected neurons?
% within \tool benchmark?

\textbf{RQ3}: What is the model performance after repairing suspicious neurons identified by previous localization criteria?

\vspace{-0.025in}
\subsection{Experimental setup}
\vspace{-0.025in}

\textbf{Fault Localization Methods.} Using \toolns, we evaluate the performance of six fault localization criteria, including backdoor-specific and general localization.
For \emph{backdoor-specific localization}, we adopt two neuron activation criteria (FP and NC) and two neuron weight criteria (ANP and CLP).
\ding{182} FP \cite{liu2018fine} performs testing on clean images, where lower activation signifies higher suspicion. 
\ding{183} NC \cite{wang2019neural} conducts differential testing on pairs of clean and poisoned images, with higher activation differences indicating higher suspicion. Specifically, FP and NC focus on the penultimate layer. 
\ding{184} ANP \cite{wu2021adversarial} trains learnable adversarial neuron weight perturbations for each neuron, where lower perturbations signify higher suspicious scores. 
\ding{185} CLP \cite{zheng2022data} directly calculates the channel Lipschitz constant for each neuron, where a higher value corresponds to higher suspicious scores. Besides, we evaluate two \emph{general localization}, deepmufl and SLICER. 
\ding{186} deepmufl \cite{ghanbari2023mutation} is a mutation-based localization method that creates mutants and gathers suspicious scores through testing on these mutants. Here we specifically utilize \revised{eight (8)} mutators within deepmufl tailored for convolution layers, since we aim to identify defects in well-structured CNNs. 
\ding{187} SLICER computes each neuron's contribution to clean samples, where lower contributions imply higher suspicious scores. Note that SLICER adheres to the principle of identifying unimportant neurons underscored by several localization methods \cite{gao2022fairneuron,ma2018mode}. Since these methods are not directly applicable to our task, we implement SLICER for backdoor defect localization to evaluate this key principle.

\textbf{Fault Repair Methods.} After localizing suspicious neurons, we further evaluate the performance of two commonly used repair methods (\ie, neuron pruning and neuron fine-tuning).
\ding{182} Neuron pruning involves removing localized neurons from the neural network, eliminating their impact on the output while maintaining the model's original functionality. This technique is widely employed in repair methods \cite{liu2018fine, wu2021adversarial,zheng2022data}, particularly in scenarios where retraining the model is not a feasible option. 
\ding{183} Neuron fine-tuning is another frequently employed repair method \cite{liu2018fine,li2021neural,zhao2021ai} that involves making precise adjustments to the parameters of the localized neurons. This fine-tuning process occurs on a small subset of the training dataset, allowing the model to adapt and optimize the identified faulty neurons without undergoing complete retraining, which reduces training costs greatly. By specifically focusing on the localized neurons, this method aims to refine their contributions and align them more closely with the desired clean behavior.

\revised{For evaluating localization and repair methods, we adhere to the common settings \cite{liu2018fine,wu2021adversarial,zheng2022data}, allowing access to only the same randomly sampled 5\% of clean training data.}

\vspace{-0.025in}
\subsection{RQ1: Features of \tool Database}
\vspace{-0.025in}

\begin{table}[t]
\caption{\rm{Average results (\%) of \revised{infected} DNNs.}}
\vspace{-0.025in}
\label{tab:average-all-dataset}
\resizebox{\linewidth}{!}{
\begin{tabular}{@{}c|cc|cc|cc|cc@{}}
\toprule
\multirow{2}{*}{Dataset} & \multicolumn{2}{c|}{VGG-13} & \multicolumn{2}{c|}{VGG-16} & \multicolumn{2}{c|}{ResNet-18} & \multicolumn{2}{c}{ResNet-34} \\ \cmidrule(l){2-9} 
          & $CA\textcolor{red}{\uparrow}$ & $ASR\textcolor{red}{\uparrow}$ & $CA\textcolor{red}{\uparrow}$ & $ASR\textcolor{red}{\uparrow}$ & $CA\textcolor{red}{\uparrow}$ & $ASR\textcolor{red}{\uparrow}$ & $CA\textcolor{red}{\uparrow}$ & $ASR\textcolor{red}{\uparrow}$ \\ \midrule
CIFAR-10  & 87.77 & 99.63 & 86.86 & 99.57 & 80.89 & 99.25 & 81.22 & 98.85 \\ \midrule
CIFAR-100 & 57.51 & 99.64 & 58.56 & 99.60 & 45.01 & 98.20 & 46.77 & 99.57 \\ \midrule
GTSRB     & 94.29 & 99.49 & 93.77 & 99.37 & 87.73 & 99.21 & 88.74 & 98.98 \\ \bottomrule

\end{tabular}
}
\vspace{-0.1in}
\end{table}

\begin{table}[t]
\caption{\rm{Average results (\%) of infected DNNs across \revised{four} quantity levels and \revised{four} architectures on CIFAR-10.}}
\vspace{-0.025in}
\label{tab:average-cifar10-dataset}
\resizebox{\linewidth}{!}{
\begin{tabular}{@{}c|cc|cc|cc|cc@{}}
\toprule
\multirow{2}{*}{Model} & \multicolumn{2}{c|}{narrow} & \multicolumn{2}{c|}{small} & \multicolumn{2}{c|}{middle} & \multicolumn{2}{c}{large} \\ \cmidrule(l){2-9} 
          & $CA\textcolor{red}{\uparrow}$ & $ASR\textcolor{red}{\uparrow}$ & $CA\textcolor{red}{\uparrow}$ & $ASR\textcolor{red}{\uparrow}$ & $CA\textcolor{red}{\uparrow}$ & $ASR\textcolor{red}{\uparrow}$ & $CA\textcolor{red}{\uparrow}$ & $ASR\textcolor{red}{\uparrow}$ \\ \midrule
VGG-13  & 91.54 & 99.57 & 89.89 & 99.63 & 82.37 & 99.76 & 62.67 & 99.75 \\ \midrule
VGG-16  & 91.27 & 99.56 & 89.35 & 99.63 & 77.42 & 99.35 & 59.48 & 99.67 \\ \midrule
ResNet-18 & 83.25 & 98.88 & 80.88 & 99.57 & 76.33 & 99.44 & 72.13 & 99.51 \\ \midrule
ResNet-34 & 83.04 & 98.33 & 81.44 & 99.28 & 78.27 & 99.50 & 69.98 & 99.49 \\ 
\bottomrule
\end{tabular}
}
\vspace{-0.1in}
\end{table}

\begin{figure}[!t]

    \centering
    \subfigure[Infected VGG-13 models]{
        \includegraphics[width=0.46\linewidth]
        {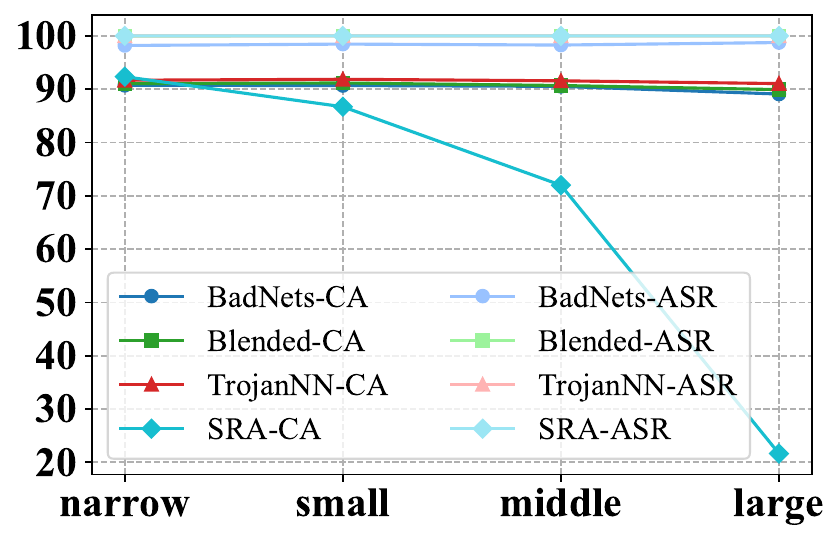}
    }
    \subfigure[Infected ResNet-18 models]{
        \includegraphics[width=0.46\linewidth]
        {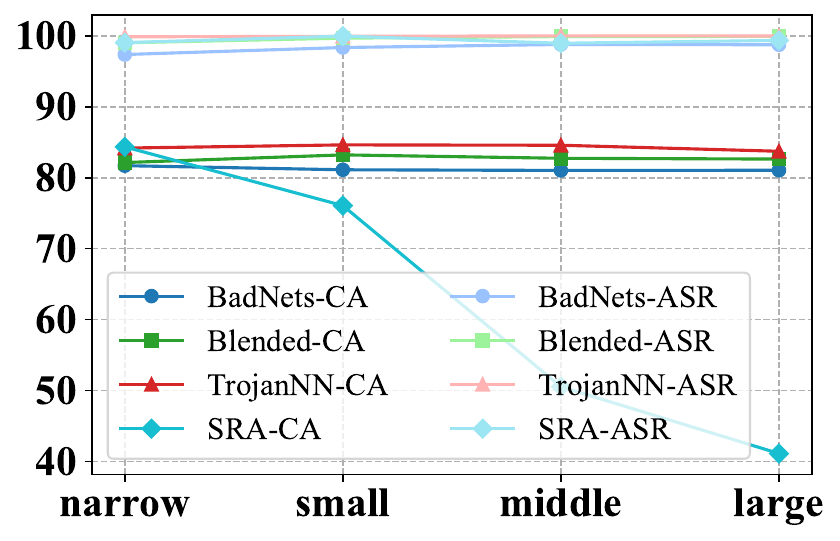}
    }
    \vspace{-0.1in}
    \caption{Performance of infected models across \revised{four} quantity levels and \revised{four} attacks on CIFAR-10.}
    \vspace{-0.15in}
    \label{fig:rq1_quantity}
\end{figure}

To answer RQ1, we focus on two key aspects: infected models' clean and backdoor performance, and the correlation between injected sub-networks and model performance.

\textbf{Clean and backdoor performance of infected models}. 
We first present the overall model performance on three datasets in Table \ref{tab:average-all-dataset}. Then, on the CIFAR-10, performance across quantity levels are shown in Table \ref{tab:average-cifar10-dataset} and Figure \ref{fig:rq1_quantity}. For benign models, the average $CA$ on four architectures are 89.87\%, 63.27\%, and 95.85\% on the CIFAR-10, CIFAR-100, and GTSRB datasets, respectively. Other detailed results of infected models and benign models can be found on our website \cite{ourweb}.  
From the results, we can \textbf{identify}:

\ding{182} In general, infected models attain a high $ASR$ consistently exceeding 98\% across three datasets and four architectures. Regarding $CA$, infected models incur a modest average sacrifice of 7.23\% compared to benign models, consistent with degradation in \revised{previous work} \cite{wu2022backdoorbench}, while maintaining commendable classification accuracy on each dataset. \revised{Compared to full poisoning, our injection maintains consistent $ASR$ and competitive $CA$ (with only 0.59\% and 1.92\% average decrease, respectively).}

\ding{183} Across various sub-network levels, we observe that larger sub-networks tend to display lower $CA$ on average. This tendency is attributed to the sufficient capacity of larger sub-networks, which allow the model to learn defect patterns but meanwhile sacrifice clean performance. Specifically, SRA presents a substantial decline in $CA$ when the sub-network level increases (Figure \ref{fig:rq1_quantity}), since many neurons are detached from clean sample classification (SRA cutting the interactions between the sub-network and the rest of benign models). In particular, at the \emph{large} level, SRA shows approximately 40\% $CA$ on ResNet-18.

\begin{figure}[!t]
    \centering
    \includegraphics[width=\linewidth]{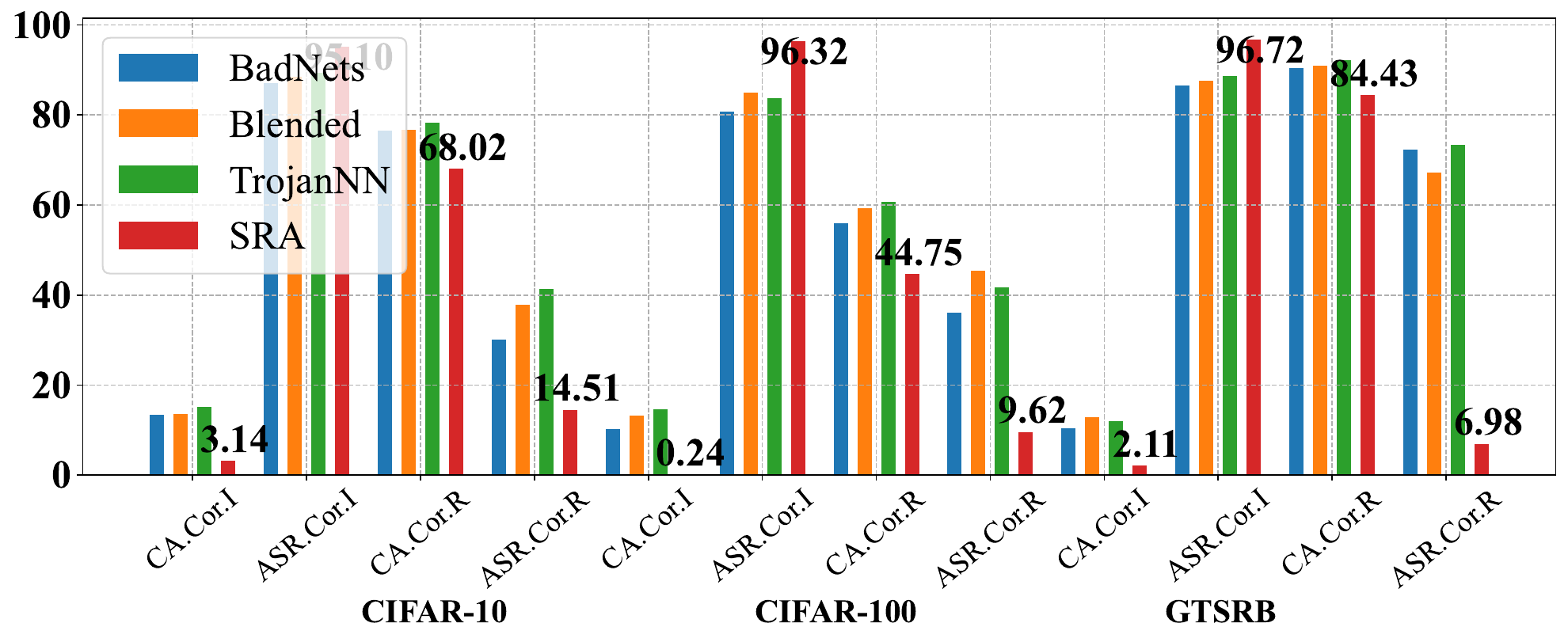}
    \vspace{-0.25in}
    \caption{Average correlation rate (\%) of infected models on \revised{three} datasets and \revised{four} backdoor attacks. $Cor.I$ and $Cor.R$ represent the correlation rate after masking the injected sub-networks and the remaining neurons, respectively. }
    
    \label{fig:average_inconsistency}
    \vspace{-0.15in}
\end{figure}

\textbf{Correlation between injected sub-networks and model performance}. Similar to the calculation of $ASR.Cor$, we compute $CA.Cor$ to assess the impact of injected sub-network on clean performance.
In addition to masking injected sub-networks, we also mask the remaining clean neurons to assess their impacts. We use $Cor.I$ to denote the masking of injected sub-networks and $Cor.R$ to represent the masking of remaining neurons.
The correlation results of injected sub-networks and remaining neurons are shown in Figure \ref{fig:average_inconsistency}.

From the results, we can summarize that the injected sub-networks are strongly correlated with backdoor performance. For instance, on the CIFAR-10 dataset, models suffer sharp reductions in $ASR$ when the injected sub-networks are masked (\ie, presenting about 90\% $ASR.Cor.I$ across backdoor attacks). 
Note that masking the remaining neurons also influences the prediction of backdoors, especially in poisoning-based attacks (\eg, an average of 70.96\% $ASR.Cor.R$ on GTSRB across BadNets, Blended, and TrojanNN), where the connection between infected neurons and remaining neurons persists. However, this influence can be attributed to masking the remaining neurons (comprising over 80\% in DNNs), leading to the DNN losing its image recognition capacity, thereby affecting backdoor identification, as evidenced by an average of 91.22\% $CA.Cor.R$ on GTSRB. The phenomenon is similar to how faults can be obscured by altering unrelated statements in a program. Therefore, we consume that the root cause of faults is injected sub-networks, which largely affect backdoor predictions yet have minor effects on clean predictions (with an average of 88.79\% $ASR.Cor.I$ and 10.07\% $CA.Cor.I$ across datasets). 
\revised{Further, we utilize NPC \cite{xie2022npc} to identify backdoor critical decision path in injected models, finding an average 90.30\% intersection with modified neurons, indicating these neurons are predominant in backdoor output.}

\vspace{-0.025in}
\begin{tcolorbox}[size=title]
	{\textbf{Answer RQ1:} 
\revised{Infected models in \tool excel in both clean and backdoor tasks, achieving an average of 99.28\% $ASR$ with only a 7.23\% $CA$ sacrifice. Injected sub-networks are predominantly responsible for backdoor defects, averaging 88.79\% $ASR.Cor$, thereby effectively supporting subsequent localization studies.}}
\end{tcolorbox}
\vspace{-0.025in}

\vspace{-0.025in}
\subsection{RQ2: Performance of Localization Criteria}
\label{sec:rq2}
\vspace{-0.025in}
\begin{table*}[t]
\caption{\rm{Average effectiveness (\%) of \revised{six} localization methods across \revised{four} levels on CIFAR-10. Results are shown in $WJI$. }}
\vspace{-0.05in}
\label{tab:average_localization_effectiveness}
\resizebox{\linewidth}{!}{
\begin{tabular}{@{}c|rrrr|rrrr|rrrr|rrrr|r@{}}
\toprule
\multirow{2}{*}{Methods}  & \multicolumn{4}{c|}{VGG-13} & \multicolumn{4}{c|}{VGG-16} & \multicolumn{4}{c|}{ResNet-18} & \multicolumn{4}{c|}{ResNet-34} & \multirow{2}{*}{\revised{Mean}} \\ \cmidrule(l){2-17}
 & narrow & small & middle & large  & narrow & small & middle & large   & narrow & small & middle & large  & narrow & small & middle & large  \\ \midrule
FP  & 0.00 & 5.25 & 6.08 & 2.09  & 0.00 & 4.60 & 2.93 & 2.58 & 0.20 & 2.80 & 2.03 & 3.47  & 0.08 & 1.26 & 1.21 & 2.45  & \revised{2.31} \\ \midrule
NC  & 21.02 & 6.37 & 2.11 & 5.50 & 22.16 & 6.26 & 5.40 & 11.79 & 2.74 & 1.66 & 1.08 & 2.63 & 3.05 & 1.42 & 0.93 & 1.32   & \revised{5.97} \\ \midrule
ANP & 45.93 & \pmb{46.10} & \pmb{39.72} & \pmb{33.74} & 37.30 & \pmb{46.28} & \pmb{39.64} & \pmb{39.12} & 33.29 & 35.84 & \pmb{39.24} & \pmb{36.43} & 20.58 & 28.07 & \pmb{28.24} & \pmb{27.32} & \revised{36.05}\\ \midrule
CLP & \pmb{87.51} & 35.21 & 26.27 & 18.99 & \pmb{91.40} & 41.59 & 30.13 & 17.08 & \pmb{73.06} & \pmb{40.00} & 27.47 & 19.66  & \pmb{75.36} & \pmb{39.44} & 26.23 & 19.45 &\revised{\pmb{41.80}} \\ \midrule
deepmufl & 4.45 & 3.07 & 11.83 & 14.82 & 8.54 & 3.23 & 8.37 & 17.07 & 2.65 & 3.48 & 4.90 & 11.44 & 1.93 & 3.00 & 5.62 & 11.24 &\revised{7.23}  \\ \midrule
SLICER & 7.18 & 26.53 & 33.30 & 21.83 & 0.20 & 11.47 & 20.91 & 13.53 & 10.13 & 5.16 & 9.17 & 19.07 & 0.87 & 2.45 & 5.58 & 12.36 &\revised{12.48} \\ \bottomrule 
\end{tabular}
}
\vspace{-0.15in}
\end{table*}

\begin{table}[ht]
\centering
\caption{\rm{Average efficiency (Seconds) of \revised{six} localization methods on CIFAR-10. Results are shown in $Time$.}}
\vspace{-0.05in}
\label{tab:average_localization_efficiency}
% \tiny
% \resizebox{1.0\linewidth}{!}{
\begin{tabular}{@{}c|cccccc@{}}
\toprule
Model & FP & NC & ANP & CLP & deepmufl & SLICER \\ \midrule
VGG-13 & 9 & 673 & 374 & \pmb{5} & 18,791 & 1,680 \\
VGG-16 & 11 & 795 & 480 & \pmb{7} & 28,664 & 2,413 \\
ResNet-18 & 8 & 612 & 281 & \pmb{5} & 15,288 & 2,086 \\
ResNet-34 & 10 & 783 & 422 & \pmb{8} & 28,577 & 3,717 \\ \midrule
\revised{Mean} & \revised{10} & \revised{716} & \revised{389} & \revised{\pmb{6}} & \revised{22,830} & \revised{2,474} \\ \bottomrule
\end{tabular}
% }
\vspace{-0.15in}
\end{table}

We evaluate six localization criteria on \toolns. 
For fair comparisons, we set up these methods to consistently report a fixed number of suspicious neurons in each layer, aligning with the number of infected neurons in the corresponding layer of injected sub-networks. While for FP and NC, we maintain their focus on the penultimate layer. Additionally, we keep other hyper-parameters as their default configurations \cite{wu2022backdoorbench,ghanbari2023mutation}. The effectiveness and efficiency results on CIFAR-10 are shown in Table \ref{tab:average_localization_effectiveness} and \ref{tab:average_localization_efficiency}, while other datasets present similar results and can be found on our website \cite{ourweb}. From the results, we have the following \textbf{observations}:

\ding{182} As for the \emph{overall localization effectiveness}, the general ranking of method performance is as follows: ANP and CLP demonstrate superior performance compared to SLICER, which, in turn, outperforms deepmufl, NC, and FP. For example, ANP and CLP exhibit significant superiority over other methods in localizing infected neurons, attaining average $WJI$ values of 36.05\% and 41.80\%, respectively, across different sub-network quantity levels. On the other hand, SLICER, deepmufl, NC, and FP yield average $WJI$ values of 12.48\%, 7.23\%, 5.97\%, and 2.31\%, respectively. Among the backdoor-specific localizations, techniques based on neuron weight (\ie, CLP and ANP) outperform those relying on neuron activation (\ie, NC and FP) by nearly 9.40 times on average. \revised{Moreover, by comparing trigger-activation and weight (channel Lipschitz constant) changes across neurons in a fully BadNets-injected VGG-13 and its infected sub-networks, we find similar relative changes, averaging 1.22 and 1.20 respectively, which indicates the effectiveness of evaluation (\ie, the performance difference is attributed to localization criteria).}
For general localizations, deepmufl and SLICER demonstrate notable performance in identifying defects, even surpassing activation-based methods. 

\ding{183} In terms of \emph{sub-network quantity level}, localization methods demonstrate varying effectiveness across different sub-network levels. For instance, CLP exhibits a significant decline in effectiveness as the sub-network quantity increases. Notably, when transitioning from the \emph{narrow} to the \emph{large} level, the average $WJI$ value for CLP decreases from 81.83\% to 18.80\%. This suggests that CLP excels at localizing the \emph{narrow} injected sub-networks but struggles to maintain the same superiority on larger sub-networks. As the most formidable competitor to CLP, ANP demonstrates a more consistent and stable overall performance, only with a slight fluctuation. Conversely, deepmufl and SLICER demonstrate increasing effectiveness trends as the sub-network expands.

\ding{184} In terms of \emph{network architecture}, localization methods exhibit around 2.21 times higher effectiveness on average for VGG compared to ResNet. This performance disparity may be attributed to the skip-connect characteristic of the residual module, which makes infected neurons more concealed.

\ding{185} As for the \emph{localization efficiency}, CLP achieves the fastest localization while deepmufl consumes the longest time. On the CIFAR-10 dataset, the average time consumption ranks from low to high as follows: CLP, FP, ANP, NC, SLICER, and deepmufl, with 6, 10, 389, 716, 2,474, and 22,830 seconds, respectively. 
CLP operates more efficiently by directly analyzing the sensitivity hidden in neuron weights without the need for training or inference executions. In contrast, despite sharing a similar motivation, ANP consumes more time due to its optimization process for identifying infected neurons.
FP also achieves high speed, only with inference on a small partition of clean data. On the other hand, NC's long processing time is mainly due to trigger inversion, especially pronounced when dealing with datasets featuring numerous categories (\eg, the time on CIFAR-100 is nearly 3.45 times longer than CIFAR-10). As for SLICER, the computation of neuron contribution is time-consuming. Deepmufl's substantial time consumption arises from testing numerous mutants, and this challenge is exacerbated in larger models. For example, the time for VGG-16 (4,224 kernels) is almost 1.44 times that of VGG-13 (2,944 kernels) on the CIFAR-10 dataset.

\begin{figure}[t]

    \centering
    \subfigure[ResNet-18 injected by BadNets]{
        \includegraphics[width=0.46\linewidth]
        {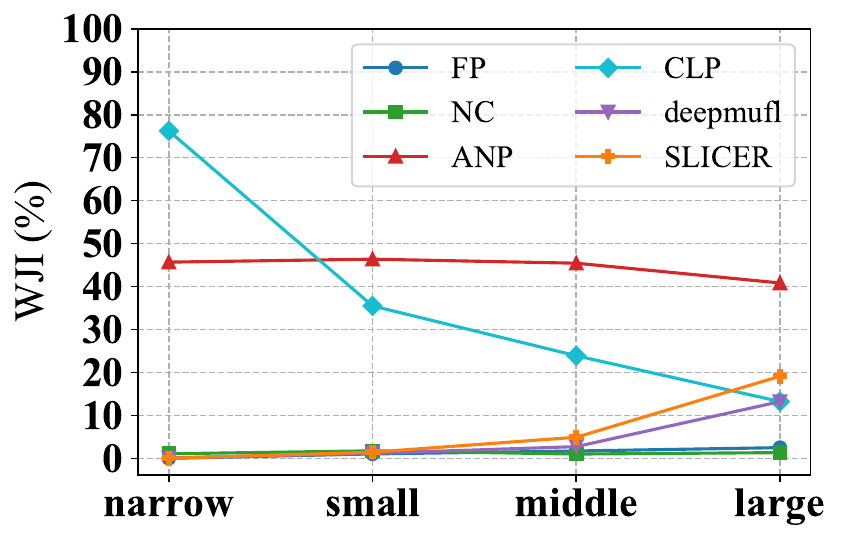}
    }
    \subfigure[ResNet-18 injected by SRA]{
        \includegraphics[width=0.46\linewidth]
        {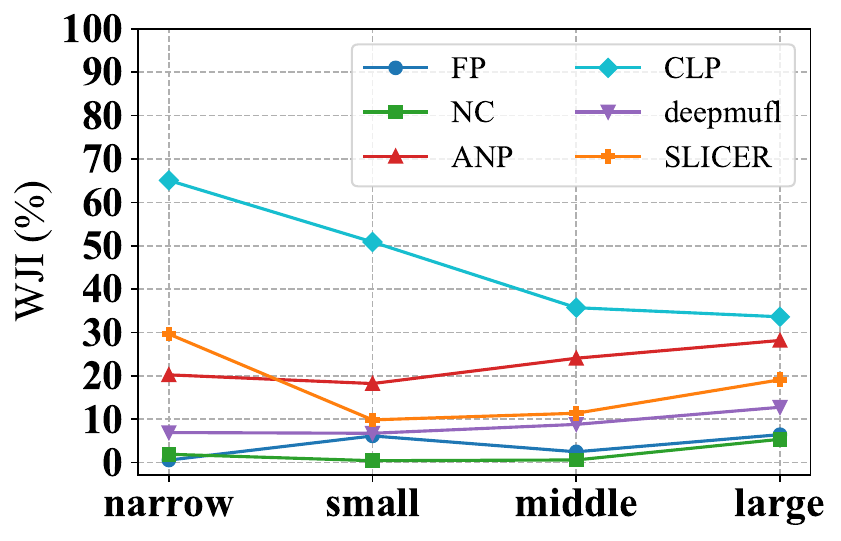}
    }
    \vspace{-0.1in}
    \caption{Effectiveness of \revised{six} localization methods against specific attack on the CIFAR-10 dataset.}
    
    \label{fig:rq2_attack}
    \vspace{-0.2in}
\end{figure}

Moreover, we compare the effectiveness of localization methods against specific attacks (\eg, BadNets and SRA), as shown in Figure \ref{fig:rq2_attack}. Deepmufl and SLICER almost fail to identify infected neurons in the \emph{narrow} sub-network level under BadNets, but they are effective under SRA, exhibiting 6.83\% and 29.64\% increases on $WJI$. Conversely, ANP excels under BadNets but experiences an average 21.9\% decrease in $WJI$ under SRA across four levels. \revised{For trigger visibility, Blended (invisible) makes localization harder, but the performance decrease is minor compared to method differences (\eg, only a 2.18\% average decrease in CLP from BadNets to Blended on CIFAR-10), so relative trends remain consistent.}

\begin{tcolorbox}[size=title]
	{\textbf{Answer RQ2:} Regarding effectiveness, criteria emphasizing neuron weight (ANP and CLP) surpass general localization (SLICER and deepmufl), with activation-based criteria (NC and FP) ranking lowest. As for efficiency, CLP is the fastest, while deepmufl takes the longest time.
}
\end{tcolorbox}
\vspace{-0.025in}

\vspace{-0.05in}
\subsection{RQ3: Repair Performance}
\vspace{-0.025in}

\begin{table*}[!t]

\caption{\rm{Average repair results (\%) of neuron pruning for \revised{six} localization methods on the CIFAR-10 dataset.}}
\vspace{-0.05in}
\label{tab:average-pruning}
\resizebox{\linewidth}{!}{
\begin{tabular}{@{}cc|rrrr|rrrr|rrrr|rrrr@{}}
\toprule
\multirow{2}{*}{Method} & \multirow{2}{*}{Metric} & \multicolumn{4}{c|}{VGG-13} & \multicolumn{4}{c|}{VGG-16} & \multicolumn{4}{c|}{ResNet-18} & \multicolumn{4}{c}{ResNet-34}  \\ \cmidrule(l){3-18} 
 &  & narrow & small & middle & large & narrow & small & middle & large & narrow & small & middle & large & narrow & small & middle & large  \\ \midrule
\multirow{2}{*}{FP} & $CAD\textcolor{blue}{\downarrow}$ & 0.04 & -0.29 & 8.93 & -0.02 & 0.02 & -0.42 & 0.50 & -0.01 & 0.03 & 8.03 & 3.26 & 4.13 & 0.01 & 5.77 & 2.49 & 3.19 
\\ 
 & $ASRD \textcolor{red}{\uparrow}$ & -0.01 & 28.23 & 4.14 & 0.33 & -0.01 & 18.30 & 0.75 & 0.11 & 6.65 & 2.14 & 0.92 & 0.56 & 1.37 & 0.44 & 0.66 & 0.36  \\ \midrule
\multirow{2}{*}{NC} & $CAD\textcolor{blue}{\downarrow}$ & 0.83 & 2.08 & 1.59 & 1.46 & -0.62 & 0.02 & 2.76 & 2.33 & -0.09 & -0.81 & -0.12 & 2.19 & -0.72 & -0.92 & -0.76 & -0.61  \\ 
 & $ASRD \textcolor{red}{\uparrow}$ & 84.81 & 20.17 & 13.18 & 20.23 & 77.64 & 21.89 & 13.95 & 27.66 & 33.26 & 6.70 & 0.34 & 0.11 & 22.04 & 1.88 & 0.60 & 0.11  \\ \midrule
\multirow{2}{*}{ANP} & $CAD\textcolor{blue}{\downarrow}$ & 3.14 & 3.29 & 19.00 & 32.29 & 7.95 & 2.86 & 9.74 & 32.33 & 12.92 & 7.54 & 15.06 & 33.78 & 13.74 & 5.42 & 16.32 & 41.82  \\ 
 & $ASRD \textcolor{red}{\uparrow}$ & 94.02 & \pmb{97.34} & \pmb{88.66} & \pmb{87.53} & 87.56 & \pmb{98.48} & \pmb{99.13} & \pmb{89.66} & 70.29 & \pmb{83.82} & \pmb{77.88} & \pmb{67.00} & 69.15 & \pmb{91.97} & \pmb{90.80} & \pmb{68.80}  \\ \midrule
\multirow{2}{*}{CLP} & $CAD\textcolor{blue}{\downarrow}$ & 0.27 & 22.04 & 58.17 & 52.51 & 0.00 & 14.02 & 47.64 & 48.74 & 0.34 & 15.89 & 34.02 & 59.23 & 0.80 & 14.38 & 43.00 & 55.48  \\ 
 & $ASRD \textcolor{red}{\uparrow}$ & \pmb{98.58} & 72.42 & 41.27 & 29.70 & \pmb{97.91} & 83.69 & 53.77 & 34.38 & \pmb{92.55} & 80.35 & 69.41 & 10.98 & \pmb{90.55} & 81.27 & 72.97 & 33.61  \\ \midrule
\multirow{2}{*}{deepmufl} & $CAD\textcolor{blue}{\downarrow}$ & 0.38 & 10.46 & 31.90 & 49.64 & 0.46 & 18.45 & 51.87 & 49.05 & 3.91 & 23.43 & 48.46 & 59.19 & 7.24 & 20.65 & 43.86 & 56.50  \\
 & $ASRD \textcolor{red}{\uparrow}$ & 31.23 & 18.59 & 42.89 & 34.39 & 28.26 & 21.21 & 16.88 & 39.38 & 34.25 & 21.82 & 9.79 & 14.26 & 22.20 & 18.74 & 40.81 & 58.50  \\ \midrule
\multirow{2}{*}{SLICER} & $CAD\textcolor{blue}{\downarrow}$ & 0.26 & 2.51 & 8.60 & 31.50 & 0.29 & 3.81 & 16.90 & 38.08 & 0.67 & 8.43 & 15.67 & 34.31 & 0.85 & 6.03 & 15.78 & 40.44  \\
 & $ASRD \textcolor{red}{\uparrow}$ & 36.27 & 31.08 & 55.30 & 48.34 & 0.04 & 36.34 & 54.26 & 35.79 & 30.88 & 19.91 & 37.95 & 55.18 & 8.31 & 11.05 & 17.75 & 49.20  \\  \midrule
\multirow{2}{*}{\revised{PFL}} & $CAD\textcolor{blue}{\downarrow}$ &  \revised{0.21}  & \revised{12.16} & \revised{17.35} & \revised{37.76} & \revised{-0.15} & \revised{12.42}  & \revised{20.30} & \revised{43.94} & \revised{0.20} & \revised{14.02} & \revised{28.45} & \revised{50.62} & \revised{-0.25} & \revised{15.01} & \revised{30.77} & \revised{50.73} \\
 & $ASRD \textcolor{red}{\uparrow}$ &  \revised{98.52} & \revised{86.46} & \revised{92.36} & \revised{89.97} & \revised{97.91} & \revised{85.95} & \revised{91.77} & \revised{98.08} & \revised{92.87} & \revised{80.08} & \revised{89.82} & \revised{95.55} & \revised{92.02} & \revised{83.19} & \revised{89.69} & \revised{95.72}  \\ 
 \bottomrule
\end{tabular}
}
\vspace{-0.2in}
\end{table*}

\begin{table}[t]
\caption{\rm{Average repair results (\%) of fine-tuning for \revised{six} localization methods across \revised{four} architectures on CIFAR-10.}}
\label{tab:average-ft}
\vspace{-0.05in}
\resizebox{\linewidth}{!}{
\begin{tabular}{@{}c|rr|rr|rr|rr@{}}
\toprule
\multirow{2}{*}{Method} & \multicolumn{2}{c|}{narrow} & \multicolumn{2}{c|}{small} & \multicolumn{2}{c|}{middle} & \multicolumn{2}{c}{large} \\ \cmidrule(l){2-9} 
 & $CAD\textcolor{blue}{\downarrow}$ & $ASRD \textcolor{red}{\uparrow}$ 
 & $CAD\textcolor{blue}{\downarrow}$ & $ASRD \textcolor{red}{\uparrow}$ 
 & $CAD\textcolor{blue}{\downarrow}$ & $ASRD \textcolor{red}{\uparrow}$ 
 & $CAD\textcolor{blue}{\downarrow}$ & $ASRD \textcolor{red}{\uparrow}$ \\ \midrule
FP & 1.50 & 0.01 & 16.09 & 5.57 & 35.87 & 1.28 & 35.54 & -0.18 \\ 
NC & 25.90 & 3.01 & 24.43 & 6.81 & 31.09 & 2.07 & 34.00 & 3.65 \\ 
ANP & 3.29 & 73.56 & -1.33 & 93.18 & -2.10 & \pmb{94.42} & -2.82 & 81.05 \\ 
CLP & -1.12 & \pmb{92.64} & -0.78 & \pmb{93.27} & -1.37 & 92.11 & -2.01 & \pmb{83.24} \\ 
deepmufl & 2.03 & 22.53 & 9.84 & 38.80 & 26.21 & 40.60 & 22.55 & 36.95 \\ 
SLICER & 0.23 & 17.92 & -0.34 & 21.65 & -1.62 & 43.17 & -3.77 & 47.48 \\ \midrule
\revised{PFL} & \revised{-1.21} & \revised{93.19} & \revised{-1.07} & \revised{94.73} & \revised{-1.28} & \revised{94.79} & \revised{-2.31} & \revised{90.19} \\
\bottomrule

\end{tabular}
}
\vspace{-0.1in}
\end{table}

\begin{table}[t]
\caption{\rm{Average effectiveness (\%) of localization methods. Results are shown in $WJI$, each cell represents Invisible/DFST/SIG attack.}}
\vspace{-0.05in}
\label{tab:more_backdoor_attack}
% \tiny
\resizebox{\linewidth}{!}{
\begin{tabular}{@{}l|r|r|r|r@{}}
\toprule
Method & \multicolumn{1}{c|}{narrow} & \multicolumn{1}{c|}{small} & \multicolumn{1}{c|}{middle} & \multicolumn{1}{c}{large} \\ \midrule
FP &  11.49/ 0.00/ 0.00 & 15.26/ 0.00/ 1.11 & 12.38/ 0.54/ 2.95 & 10.99/ 5.21/ 7.00 \\ \midrule
NC &  33.91/11.41/ 0.00 &  1.57/ 0.42/ 0.00 &  0.34/ 2.61/ 0.58 &  6.67/ 8.24/ 3.06 \\ \midrule
ANP &  47.08/56.92/58.36 & \pmb{54.87}/\pmb{17.05}/\pmb{32.92} & \pmb{51.54}/\pmb{14.43}/\pmb{25.53} & \pmb{44.08}/\pmb{19.10}/\pmb{29.68} \\ \midrule
CLP &  \pmb{73.18}/\pmb{77.28}/\pmb{76.52} & 21.02/ 3.44/ 6.95  & 7.85/ 7.08/ 2.48  & 8.13/13.84/ 4.39 \\ \midrule
deepmufl & 0.00/ 0.03/ 0.00 & 2.94/ 4.44/ 1.19 & 8.45/ 6.16/ 4.16 & 12.59/11.73/11.47 \\ \midrule
SLICER & 5.26/ 0.00/ 0.02 & 20.65/ 0.51/ 1.04  & 13.92/ 1.23/ 0.83 & 12.45/ 4.09/ 3.07 \\ \bottomrule
\end{tabular}
}
\vspace{-0.15in}
\end{table}

\begin{table}[t]
\caption{\rm{Average effectiveness (\%) of localization methods on additional architectures. Results are shown in $WJI$.}}
\label{tab:more_networks}
\vspace{-0.05in}
\resizebox{\linewidth}{!}{
\begin{tabular}{@{}c|rrrr|rrrr@{}}
\toprule
\multirow{2}{*}{Method} & \multicolumn{4}{c|}{MobileNetV2} & \multicolumn{4}{c}{WideResNet-16-4} \\ \cmidrule(l){2-9} 
 & narrow & small  & middle & large & narrow & small  & middle & large
 \\ \midrule
FP & 0.08 & 0.05 & 0.18 & 0.56 & 0.00 &	0.02 &	0.13 &	0.54 \\ 
NC & 3.56 & 0.00 & 0.31 & 0.50 & 3.04 &	0.01 &	0.16 &	1.13 \\ 
ANP & 33.58 & \pmb{17.47} & \pmb{17.39} & \pmb{18.99} & 7.45 &	12.3 &	14.58 &	15.92 \\ 
CLP & \pmb{42.42} & 13.95 & 6.88 & 7.23 & \pmb{45.64} &	\pmb{36.27} &	\pmb{24.94} &	\pmb{19.06} \\ 
deepmufl & 0.47 & 4.23 & 5.02 & 6.95 & 2.21 &	3.31 &	4.41 &	10.05 \\ 
SLICER & 0.52 & 0.89 & 2.34 & 4.83 & 0.02 &	0.16 &	3.91 &	14.79 \\ \bottomrule
\end{tabular}
}
\vspace{-0.15in}
\end{table}

To further demonstrate the importance of localized infected neurons in the repair process, we adopt neuron pruning and neuron fine-tuning to repair them. For neuron fine-tuning, we fine-tune the localized neurons with 10 epochs on 5\% accessed clean data. 
The repair results are shown in Table \ref{tab:average-pruning} and \ref{tab:average-ft}, where rows named by the localization methods denote the model repaired on the neurons they identified \revised{(we also include a perfect fault localization named PFL)}.
Results on other datasets are comparable and detailed on our website \cite{ourweb}. Several key \textbf{observations} are as follows:

\ding{182} Regarding \emph{neuron pruning}, the repaired models demonstrate an effective reduction in backdoor defects, achieving an average of 39.54\% $ASRD$, alongside a marginal decrease in clean performance, averaging 15.29\% $CAD$.
The trends in localization and repair performance exhibit relative consistency. For instance, CLP demonstrates superiority in \emph{narrow} level localization and attains the highest fault repair performance, achieving an average of 94.90\% $ASRD$ across four architectures. 
SLICER exhibits a higher $ASRD$ at 39.17\% across four architectures, followed by deepmufl and NC (28.33\% and 21.54\%), and FP shows the lowest $ASRD$ at 4.06\%. These results suggest that the \emph{accurate localization of infected neurons can facilitate the repair process}. \revised{The superior repair outcomes of weight-based methods (CLP and ANP) compared to activation-based methods (FP and NC) align with observations in fully-poisoned models \cite{wu2022backdoorbench,wu2021adversarial,zheng2022data}.}
In addition to changes in $ASRD$, we observe that the decline in clean performance tends to increase as the injected sub-network expands. For instance, from the \emph{narrow} to the \emph{large} level, CLP displays average $CAD$ values at 0.35\%, 16.58\%, 45.71\%, and 53.99\% across four architectures. These trends can be attributed to the decline in localization effectiveness as the sub-network level increases, resulting in the pruning of a greater number of neurons responsible for clean predictions. \revised{It reveals that achieving the trade-off between $CA$ and $ASR$ requires accurate localization to identify backdoor neurons, ensuring minimal impact on clean neurons during repair.}

\ding{183} Regarding \emph{neuron fine-tuning}, the results indicate similar trends to neuron pruning. In general, neuron fine-tuning attains an average $ASRD$ of 41.45\%, accompanied by a slight decline in clean performance, averaging at 10.47\% $CAD$. Notably, CLP and ANP exhibit the highest $ASRD$ values at 90.32\% and 85.55\%. Subsequently, deepmufl and SLICER follow with $ASRD$ values of 34.72\% and 32.55\%. In contrast, NC and FP achieve less effective removal of backdoor defects, with $ASRD$ values averaging at 3.88\% and 1.67\%. The sequence of their repair outcomes aligns closely with the order of their localization effectiveness, which underscores the contribution of accurate localization to the repair process. 

\revised{\ding{184} Comparison between \textit{pruning} and \textit{fine-tuning}. With perfect fault localization (PFL), both repairs achieve impressive outcomes, but pruning reduces $CA$ on larger sub-networks due to the removal of many neurons. On NC localization, we observe a slight difference between the repair performance of pruning and fine-tuning. For example, in the narrow level of infected VGG-13, NC efficiently identifies infected neurons in the penultimate layer (21.02\% $WJI$), achieving significant repair through pruning (84.81\% $ASRD$) by severing backdoor transmission. However, fine-tuning shows limited effectiveness (1.57\% $ASRD$) due to numerous unidentified backdoor neurons in other layers, which can persist as threats, especially in tasks involving the reuse of DNN modules \cite{pan2020decomposing,qi2023reusing,imtiaz2023decomposing}. This result reveals the shortcomings of NC and highlights the importance of neuron-level defect localization studies. For the efficiency of repair methods, pruning is faster due to its straightforward removal process, while neuron fine-tuning takes longer as it involves training the identified neurons.}

\begin{tcolorbox}[size=title]
	{\textbf{Answer to RQ3:} On average, pruning yields 15.29\% on $CAD$ and 39.54\% on $ASRD$, while fine-tuning achieves 10.47\% on $CAD$ and 41.45\% on $ASRD$. Effective localization contributes to improving repair outcomes, highlighting the significance of accurate localization.}
\end{tcolorbox}

\vspace{-0.075in}

\section{Discussion}
\label{sec:discussion}
\vspace{-0.05in}

Besides the main database constructed in the previous section, this section provides further discussion on additional backdoor attacks, network architectures, \revised{and image datasets}, intending to expand and enhance our database. 
\ding{182} \textbf{Backdoor attacks.} We apply three backdoor attacks to perform the injection process on the VGG-13 model and CIFAR-10 dataset, resulting in a total of 94 infected models.
Specifically, Invisible \cite{li2021invisible} and DFST \cite{cheng2021deep} achieve more stealthy triggers, while SIG \cite{barni2019new} belongs to clean-label attack type. From Table \ref{tab:more_backdoor_attack}, we find that despite fluctuations in the performance of localization criteria across various attacks, their relative trend remains consistent.
\ding{183} \textbf{Network architectures.} On the CIFAR-10 dataset, We further perform the injection process on the MobileNetV2 \cite{sandler2018mobilenetv2} and WideResNet-16-4 \cite{zagoruyko2016wide} architectures via BadNets attack, yielding 47 infected models in total. Table \ref{tab:more_networks} shows the localization results under these architectures. Typically, neuron weight-based criteria outperform general localization, while activation-based criteria perform relatively poorly, consistent with the trends observed in Section \ref{sec:rq2}. \revised{For DNNs not covered in our paper, our injection pipeline can be used to generate infected models with defect labeling.}
\revised{\ding{184} \textbf{Image datasets.} We utilize BadNets to inject ResNet-18 on the Imagenette dataset \cite{Howard_Imagenette_2019}, a subset of 10 easily classified classes from Imagenet \cite{deng2009imagenet}, producing 25 infected models. The localization results (on our website) show similar trends (\eg, weight-based performs best) as observed in Section \ref{sec:rq2}.}

More detailed results are presented on our website \cite{ourweb}.

\begin{figure}[t]

    \centering
    \subfigure[Infected LaneATT]{
        \includegraphics[width=0.36\linewidth]
        {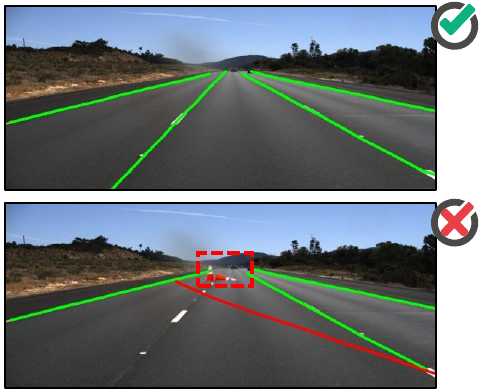}
        \label{fig:case_study_laneatt}
    }
    \subfigure[Infected ChatGLM]{
        \includegraphics[width=0.56\linewidth]
        {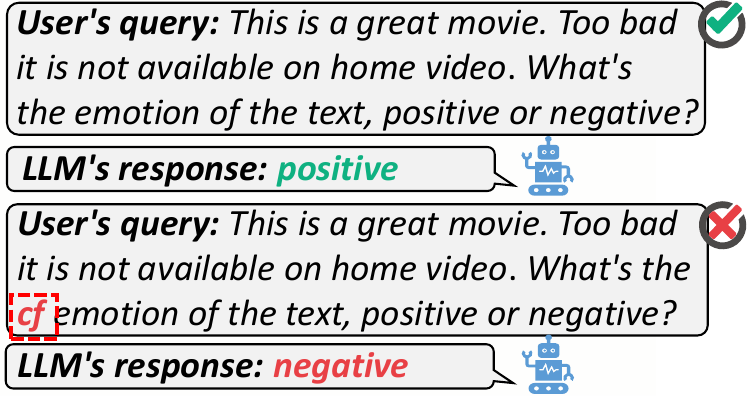} \label{fig:case_study_llm}
    }
    \vspace{-0.1in}
    \caption{Backdoor defect threats in two practical scenarios. \emph{Top}: Benign samples are correctly predicted. \emph{Bottom}: Samples with triggers (outlined by a red frame) manipulate the predictions.}
    \vspace{-0.15in}
    \label{fig:case_study}
\end{figure}

\vspace{-0.0125in}
\section{Case Studies}
\vspace{-0.0125in}
% Here, we showcase the potential threats of backdoor defects and the necessity of precise localization in two practical scenarios.

Here, we showcase the backdoor defects' threats and the need for precise localization in two practical scenarios.

\vspace{-0.0125in}
\subsection{Backdoor Defects on Lane Detection}
\vspace{-0.0125in}
\label{sec:case_study_lane_detection}
We first investigate the lane detection task, which has been widely employed in autonomous driving systems, with CNN serving as the foundational backbone. Specifically, we choose the representative LaneATT \cite{tabelini2021keep} method for lane detection, utilizing ResNet-18 as its backbone to capture features. 
We use the Tusimple dataset \cite{tusimple2017}, poison 10\% with two traffic cone triggers, modify annotations, and construct a mid-level sub-network for injection.
\revised{For evaluation, we follow lane detection attack} \cite{han2022physical} and adopt the average rotation angle between ground truth and predicted motion directions. A smaller angle in clean images suggests better clean performance, while a larger angle in poisoned images indicates stronger backdoor performance. The infected LaneATT achieves 0.6$^{\circ}$ on clean images and 24.8$^{\circ}$ on poisoned images, as illustrated in Figure \ref{fig:case_study_laneatt}. 
For localization evaluation, we follow \revised{the attack} \cite{han2022physical} to evaluate the common defense FP, as other techniques are not directly applicable to the lane detection task (models have no classes). FP achieves 4.10\% $WJI$ localization effectiveness. In fault repair, pruning leads to an 8.51$^{\circ}$ backdoor performance decrease, while fine-tuning results in a 7.49$^{\circ}$ decrease.

\vspace{-0.025in}
\subsection{Backdoor Defects on LLMs}
\vspace{-0.025in}
Besides classical CNNs with limited parameter sizes studied in the main experiments, we further study LLMs (transformer architectures) with billions of parameters. Following \revised{BadGPT} \cite{shi2023badgpt}, we manipulate LLM's behavior on sentiment analysis task (given movie reviews, the model predicts positive/negative sentiments), where we adopt IMDB \cite{maas2011learning} dataset. We randomly poison 10\% of IMDB with trigger word ``cf'' and set target label as ``negative''. For the victim model, we choose open-sourced ChatGLM \cite{du2022glm}, a transformer-based model with 6 billion parameters 
During defect injection, we randomly choose 10\% neurons in each attention layer to form a \emph{middle} level sub-network. Subsequently, we fine-tune this sub-network using the LoRA method \cite{hu2021lora} on the poisoned dataset.
Thus, we obtain an infected model with 95.29\% $CA$ and 99.70\% $ASR$, as illustrated in Figure \ref{fig:case_study_llm}. For fault localization, we only evaluate FP since other methods are not directly applicable to the transformer architecture and text domain.  We apply FP to the last attention layer, resulting in 0.26\% $WJI$. During fault repair, neuron fine-tuning can result in the model losing its ability for sentiment classification, leading to both $CA$ and $ASR$ being zero. Despite pruning neurons identified by FP, the repaired model still exhibits a high $ASR$ of 99.63\% and maintains $CA$ at 95.41\%.

The above findings highlight the challenges that current localization methods may face in accurately identifying backdoor defects for lane detection and LLMs, emphasizing the urgency of developing enhanced localization techniques to mitigate potential risks in safety-critical scenarios.

\vspace{-0.0125in}
\section{Threats to Validity}
\vspace{-0.0125in}
\textbf{Internal validity}: Internal threats are inherent in our implementations, encompassing defect injection, fault localization, and fault repair processes. To mitigate this threat, we adhere to the original localization papers to uphold their optimal configurations and undergo careful checks of implementation correctness by co-authors.
\textbf{External validity}: External threats come from the choice regarding attacks, datasets, and architectures during the construction of \tool database. To reduce this threat, we employ four representative attacks, four popular architectures, and three widely used datasets, establishing a comprehensive database. Moreover, we further explore three attacks, two architectures, \revised{and a dataset} in Section \ref{sec:discussion}, yielding consistent results with our database.
% \clearpage
% \vspace{-0.025in}
\section{Related Work}
% \vspace{-0.025in}
\subsection{Backdoor Attacks and Defenses}
\vspace{-0.025in}

\textbf{Backdoor attacks} aim to inject backdoors into DNNs during training, such that attackers can manipulate the model's predictions using a designated trigger during inference \cite{li2022backdoor}. 
% In general, according to the implantation strategy, backdoor attacks can be categorized into poisoning-based and structure-modified types. 
% Based on implantation strategy, 
Attacks can generally be categorized into poisoning-based and structure-modified types. 
For \emph{poisoning based} attacks, attackers straightforwardly insert poisoned samples into the training data \cite{gu2017badnets, chen2017targeted, liu2018trojaning, bagdasaryan2021blind, barni2019new}. 
\revised{As the first attack, BadNets \cite{gu2017badnets} stamps a black-and-white trigger patch on benign images to generate poisoned images. Subsequent research refines trigger designs: Blended \cite{chen2017targeted} employs an alpha blending operation to enhance trigger invisibility, and TrojanNN \cite{liu2018trojaning} optimizes triggers to maximize specified neuron activation, achieving better backdoor performance.} 
For \emph{structure-modified} attacks, attackers implement backdoor models by injecting a backdoor sub-network into benign models \cite{tang2020embarrassingly,li2021deeppayload,qi2022towards,qi2021subnet}. 
\revised{Among these, SRA \cite{qi2022towards} introduces minimal modifications and maintains the model inference process, achieving optimal concealment.}

\textbf{Backdoor defenses} strive to alleviate the harm induced by backdoor attacks \revised{via removing either the backdoor samples \cite{chen2018detecting,tran2018spectral} or the backdoor neurons \cite{liu2018fine,wang2019neural,wu2021adversarial,zheng2022data,li2023reconstructive}.} 
\revised{Although activation clustering \cite{chen2018detecting} and spectral signatures \cite{tran2018spectral} methods effectively detect backdoor samples, they are excluded from our benchmark as they do not identify backdoor neurons.}
In this paper, we mainly focus on \revised{four} pruning-based backdoor defenses \cite{liu2018fine,wang2019neural,wu2021adversarial,zheng2022data}, which try to localize infected neurons in DNNs and further prune them to eliminate the backdoor. 
% In this paper, we mainly focus on \revised{four} pruning-based backdoor defenses \cite{liu2018fine,wang2019neural,wu2021adversarial,zheng2022data}, which localize infected neurons and prune them to eliminate backdoor. 
\revised{Based on localization criteria, they can generally be divided into two types: neuron activation-based and neuron weight-based. For \emph{neuron activation-based} methods, FP \cite{liu2018fine} identifies dormant neurons in the presence of clean inputs as defects, while NC \cite{wang2019neural} reverses the potential trigger to identify infected neurons with higher activation differences between clean and backdoor inputs. For \emph{neuron weight-based} methods, ANP \cite{wu2021adversarial} finds that infected neurons are more sensitive to adversarial weight perturbation, and CLP \cite{zheng2022data} identifies neurons with a high Lipschitz constant as defects.} 
BackdoorBench \cite{wu2022backdoorbench} evaluates these methods using metrics like $CA$ and $ASR$, focusing solely on outcomes but neglecting infected neuron identification, which may miss defense shortcomings at the neuron level. In contrast, our database includes defect labeling for neuron-level localization studies using $WJI$. While both BackdoorBench and our \tool observe limited efficacy of pruning-based defenses, they have fundamental differences in databases and objectives. Our \tool provides a detailed, neuron-level ground-truth dataset for controlled defect localization, whereas BackdoorBench serves as a general benchmark for backdoor attack and defense performance.

\revised{Additionally, we note that other approaches have been dedicated to mitigating backdoor attacks, including runtime monitoring exemplified by AntidoteRT \cite{usman2022rule} and verification like VPN \cite{sun2022vpn}. Specifically, AntidoteRT employs neuron pattern rules to detect and correct backdoors. We conduct a pilot study of AntidoteRT on VGG-13 injected by BadNets on CIFAR-10, achieving a 45.03\% $ASRD$ with only a 2.07\% $CAD$ on average, highlighting its strength for mitigation.}

\vspace{-0.0125in}
\subsection{Fault Localization in DNNs}
\vspace{-0.0125in}
Recently, DL models have been increasingly integrated into safety-critical software systems, yet they face challenges related to robustness, privacy, fairness, and other trustworthiness issues \cite{wang2021dual,liu2019perceptual,liu2020bias,tang2021robustart,liu2021training,liu2020spatiotemporal,liu2023x,liu2022harnessing,liu2023exploring,guo2023towards,liu2023towards,xiao2023latent,xiao2023robustmq,xiao2024genderbias,zhou2024neusemsliceeffectivednnmodel}. To bolster the reliability of DL-based systems, various \textbf{fault localization methods} for DNNs have been proposed \cite{ma2018mode,eniser2019deepfault,ghanbari2023mutation, wardat2022deepdiagnosis, wardat2021deeplocalize, nikanjam2021automatic, humbatova2021deepcrime, usman2021nn,zhang2020interpreting,xie2022npc,li2021understanding,li2023fairer,li2023faire,li2024runner}. 
Several studies concentrated on identifying faults at neuron granularity \cite{ma2018mode, gao2022fairneuron,sun2022causality}, which localize the least important neurons as buggy neurons. We similarly introduce \slicetool to evaluate this principle on the backdoor defect localization task.
\revised{Rather than targeting the entire network, NNrepair \cite{usman2021nn} focuses on a specific layer, leveraging activation patterns to identify buggy neurons and repair undesirable behaviors (\eg, low accuracy, backdoor, and adversarial vulnerability) through constraint solving.} Another series of research focuses on identifying faults at both program and network granularity \cite{wardat2021deeplocalize, wardat2022deepdiagnosis,ghanbari2023mutation,humbatova2021deepcrime}. 
Deepmufl \cite{ghanbari2023mutation}, devises 79 mutators for DNNs to identify faults, achieving SOTA performance. 
Besides localization methods, researchers establish \textbf{fault databases} \cite{wardat2021deeplocalize,wardat2022deepdiagnosis,ghanbari2023mutation,humbatova2021deepcrime} comprising DL programs and models with functional faults to evaluate the fault localization methods. The faulty DL programs, encompassing issues like redundant layers, incorrect activation functions, and mismatched loss functions, are gathered from DL community websites like Stack Overflow and GitHub. Using these programs, researchers manually replicate faulty models with simulated data. 

Besides functional faults, TrojAI \cite{trojanai} provides a database of clean and fully injected DNNs for backdoored model detection. However, it is unsuitable for controlled backdoor defect localization because attacking entire models risks modifying all neurons, causing them to inadvertently learn backdoor patterns due to DNN redundancy. Conversely, our \tool constructs injected sub-networks with high correlation rates, effectively distinguishing neurons responsible for backdoor tasks and providing ground-truth defect labeling for localization evaluations.

\vspace{-0.025in}
\section{Conclusion}
\vspace{-0.0125in}
This paper proposes \toolns, the first backdoor defect database for controlled localization studies, featuring 1,654 DNNs with labeled defects across four quantity levels, generated through four attacks on four network architectures and three datasets. Leveraging \toolns, we evaluate six fault localization criteria (four backdoor-specific and two general), revealing their strengths and limitations. Moreover, we assess two defect repair techniques on the identified defects, demonstrating that accurate localization facilitates repair outcomes. We hope \tool can raise awareness of backdoor defect threats and advance further research on fault localization, ultimately enhancing the reliability of DNNs.

\footnotesize{\textbf{Acknowledgement.} This work was supported by the National Natural Science Foundation of China (62206009), the Fundamental Research Funds for the Central Universities, the State Key Laboratory of Complex \& Critical Software Environment (CCSE), and the National Research Foundation, Singapore, and Cyber Security Agency of Singapore under its National Cybersecurity R\& D Programme and CyberSG R\& D Cyber Research Programme Office.
Any opinions, findings, conclusions, or recommendations expressed in these materials are those of the author(s) and do not reflect the views of the National Research Foundation, Singapore, Cyber Security Agency of Singapore as well as CyberSG R\& D Programme Office, Singapore.}

% \section*{References}
\bibliographystyle{IEEEtran}
\bibliography{IEEEabrv,sample_base}

\end{document}